\documentclass[journal, twocolumn]{IEEEtran}
\usepackage{amsmath, amsfonts, amssymb, amsthm}
\usepackage{bbm}
\usepackage{graphicx}
\usepackage{cite}
\usepackage{mathtools}
\usepackage{subcaption}
\usepackage{color}

\newtheorem{theorem}{Theorem}
\newtheorem{lemma}{Lemma}

\newtheorem{corollary}{Corollary}
\newtheorem{definition}{Definition}
\newtheorem{remark}{Remark}

\newtheorem{assumption}{Assumption}
\newenvironment{sketch}{\proof}{\endproof}

\allowdisplaybreaks 

\begin{document}

\title{On the Price of Decentralization\\ in Decentralized Detection}

\author{Bruce (Yu-Chieh) Huang,~\IEEEmembership{Student Member,~IEEE,}~and I-Hsiang Wang,~\IEEEmembership{Member,~IEEE}
\thanks{This work was supported by NSTC of Taiwan under Grant 110-2634-F-002-029 and 111-2628-E-002-005-MY2 and NTU under Grant 112L893204. 
The material in this paper was partly presented at the 2020 IEEE Information Theory Workshop~\cite{huang2021social}.}
\thanks{B. Huang was with the Graduate Institute of Communication Engineering, National Taiwan University, Taipei, Taiwan. He is now with the Department of Electrical and Computer Engineering, University of California, Los Angeles, USA (email: brucehuang@ucla.edu).}
\thanks{I.-H. Wang is with the Department of Electrical Engineering and the Graduate Institute of Communication Engineering, National Taiwan University, Taipei, Taiwan (email: ihwang@ntu.edu.tw).}
}

\maketitle

\begin{abstract}
Fundamental limits on the error probabilities of a family of decentralized detection algorithms (eg., the social learning rule proposed by Lalitha {\it et al.}~\cite{lalitha2018social}) over directed graphs are investigated. In decentralized detection, a network of nodes locally exchanging information about the samples they observe with their neighbors to collectively infer the underlying unknown hypothesis. Each node in the network weighs the messages received from its neighbors to form its private belief and only requires knowledge of the data generating distribution of its observation. In this work, it is first shown that while the original social learning rule of Lalitha {\it et al.}~\cite{lalitha2018social} achieves asymptotically vanishing error probabilities as the number of samples tends to infinity, it suffers a gap in the achievable error exponent compared to the centralized case. The gap is due to the network imbalance caused by the local weights that each node chooses to weigh the messages received from its neighbors. To close this gap, a modified learning rule is proposed and shown to achieve error exponents as large as those in the centralized setup. This implies that there is essentially no first-order penalty caused by decentralization in the exponentially decaying rate of error probabilities. To elucidate the price of decentralization, further analysis on the higher-order asymptotics of the error probability is conducted. It turns out that the price is at most a constant multiplicative factor in the error probability, equivalent to an $o(1/t)$ additive gap in the error exponent, where $t$ is the number of samples observed by each agent in the network and the number of rounds of information exchange. This constant depends on the network connectivity and captures the level of network imbalance. Results of simulation on the error probability supporting our learning rule are shown. Further discussions and extensions of results are also presented.
\end{abstract}

\section{Introduction} \label{sec:intro}
Decentralization is one of the major themes in the development of Internet of Things (IoT), and among many different scenarios of decentralization, an important one is \emph{decentralized detection}. In decentralized detection (hypothesis testing), a group of agents (nodes) form a network (directed graph) to exchange information regarding their observed data samples in a decentralized manner, so that each of them can detect the hidden parameter that governs the sample-generating statistical model. For hypothesis testing, prior to information exchange, decentralization typically requires each node to get only full access to its samples but not the others'. In addition, each node only knows the likelihood functions of its observations.

To fulfill these requirements, a natural approach based on message passing for decentralized detection has been considered in~\cite{jadbabaie2012non, jadbabaie2013information, ShahrampourRakhlin_16, lalitha2018social, matta2019exponential}, where each node performs a local Bayesian update and sends its belief vectors (message) to its neighbors for a further consensus step. For instance, in \cite{lalitha2018social}, each node performs a consensus averaging on a re-weighting of the log-beliefs after receiving the messages (which are log-beliefs in~\cite{lalitha2018social}) from its neighbors), and the weights are summarized into a right stochastic matrix (called the ``weight matrix'', which could be viewed as the transition matrix of a Markov chain. Such an approach is termed \emph{social learning} in \cite{lalitha2018social}. Under the learning rule, it is shown that the belief on the true hypothesis converges to $1$ exponentially fast with rate characterized in~\cite{lalitha2018social} and further non-asymptotic characterization in \cite{ShahrampourRakhlin_16}. It has been noted that the concentration of beliefs depends on the network topology as well as the chosen weights.

While most literature focuses on the convergence of beliefs \cite{jadbabaie2012non, jadbabaie2013information, ShahrampourRakhlin_16, lalitha2018social, matta2019exponential}, few look into the convergence of error probability \cite{bajovic2010distributed, bajovic2012large, bajovic2016distributed}, which is arguably the most direct performance metric in hypothesis testing problems. As the convergence of error probability has not been well understood, it remains unclear what the price of decentralization on the detection performance is. There are several natural questions to be addressed. First, what is the optimal probability of error when these belief-consensus-based learning rules are utilized, and how does it depend on the network topology as well as the weights chosen by each node? Compared to the centralized performance, how much is lost? Second, with slight global knowledge about the policies of other nodes, how to improve the probability of error? Can it approach the performance of the centralized case? If it can, what is the additional cost for obtaining the needed global information?

\subsection{Contribution}
\begin{figure}[!t]
\centering
\includegraphics[width=\linewidth]{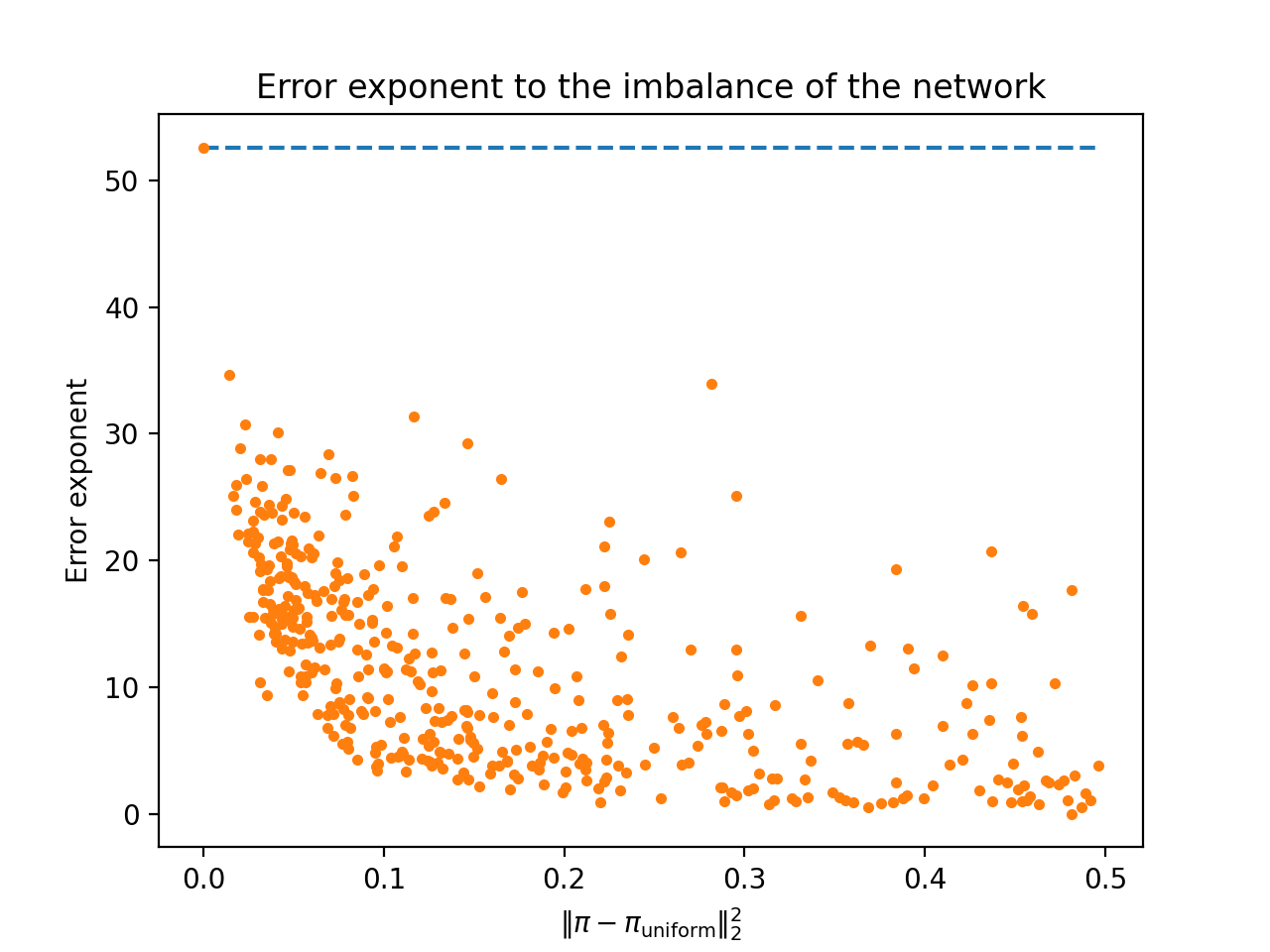}
\caption{\centering Effect of network imbalance.}
\label{fig:effect_of_imbalance}
\end{figure}

In this work, the above questions are addressed in the case of binary detection. We propose a generalization of the social learning rule in~\cite{lalitha2018social} and characterize 
the error exponents using tools in large deviation theory~\cite{dembo_zeitouni_2010}. As a result, the error exponents of the original learning rule in~\cite{lalitha2018social} are characterized, which turn out to be strictly smaller than the error exponents in the centralized case. The reason is that the decentralized sources are not weighted equally due to the convergence of the Markov chain governing the consensus. Figure~\ref{fig:effect_of_imbalance} illustrates the gap in error exponents with a simple example. In the example, 300 scale-free networks with 100 nodes in each are sampled. Each node serves as an independent Bernoulli source having consensus weights uniformly distributed to its neighbors. Gathering the consensus weights into a right stochastic matrix, the Markov chain with such corresponding transition matrix induces a unique stationary distribution denoted by $\pi$ under some minor assumptions. The figure shows that the error exponent of the original learning rule decreases with the network \emph{imbalance}. We quantify the imbalance of the network with the 2-norm between $\pi$ and the uniform stationary distribution with each entry being 0.01 for this case. Notice that only when the network is \emph{balanced}, the original learning rule obtains the optimal error exponent depicted by the blue dashed line.

The proposed generalization compensates for the imbalance of the original network consensus.
To do so, the likelihood functions in the learning rule in~\cite{lalitha2018social} are weighted \emph{geometrically} (that is, they are raised to different exponents) to equalize the importance of the sources. We show that if each agent knows the value of the stationary distribution of the consensus Markov Chain at that node, the optimal error exponent in the centralized case is achieved by properly choosing the geometric weightings. Since the first-order results do not reveal the price of decentralization, we further derive upper bounds on the higher-order asymptotics by extending Strassen's seminal result~\cite{strassen1962asymptotische} for the centralized case to our decentralized setting with the aid of the non-i.i.d. version of Esseen's theorem~\cite{esseen1945fourier, cramer2004random} and the convergence result on the Markov chains~\cite{diaconis1991geometric}. 
It turns out that the effect of decentralization is revealed as at most a constant term in the higher-order asymptotics.

The value of the stationary distribution at each node is the slight global information that enables each agent to achieve the centralized error exponent. To obtain such global knowledge, we propose a simple decentralized iterative estimation method. The estimation method only requires bi-directional communication for each pair of nodes forming a directed edge in the network. The estimation error on the stationary distribution vanishes exponentially with the number of iterations by the convergence result on Markov chains~\cite{diaconis1991geometric}. Numerical results suggest that the gap between the optimal error exponent and that with the geometric weightings being the estimated stationary distribution also vanishes exponentially with the number of iterations.

Part of the work has been published at the 2020 IEEE Information Theory Workshop~\cite{huang2021social} including Theorem~\ref{thm:decentralized_optmization_form},~\ref{thm:neyman_pearson_exponent},~\ref{thm:bayes_exponent}, and~\ref{thm:neyman_pearson_higher_order}. Additionally in this journal version, Corollary~\ref{cor:time_delay} and Theorem~\ref{thm:bayes_higher_order},~\ref{thm:gaussian} in Section~\ref{sec:higher-order} capture the constant time delay in the decentralized case and characterize the bound on the higher-order asymptotics of the Bayes risk. Furthermore, in Section~\ref{sec:simulations}, we demonstrate the impact of network imbalance, the performance of our proposed learning rule, and the effect of quantized communications. In Section~\ref{sec:discussion}, we discuss the cases where assumptions are removed and we show that our results could be extended to the case of multiple hypothesis testing.

\subsection{Related Work}
The overview papers \cite{nedic2016tutorial, sayed2013diffusion} provide extensive surveys on the algorithms and results for distributed learning. As for distributed hypothesis testing, the convergence of beliefs is considered in~\cite{lalitha2018social, jadbabaie2012non, jadbabaie2013information, ShahrampourRakhlin_16, nedic2017fast, matta2019exponential, mitra2019new, su2016asynchronous}. A learning rule adopting linear consensus on the beliefs (in contrast to the log-beliefs considered in this work) is studied in~\cite{jadbabaie2012non, jadbabaie2013information}, while~\cite{lalitha2018social} achieves a strictly larger rate of convergence by adopting consensus over the log-beliefs. An iterative local strategy for belief updates is investigated in~\cite{ShahrampourRakhlin_16}, and a non-asymptotic bound on the convergence of beliefs is provided. Based on the work in~\cite{lalitha2018social}, the convergence of beliefs is studied under the setting of weakly connected heterogeneous networks in~\cite{matta2019exponential} where the true hypothesis might be different among the components of the network.
Error exponents are studied in~\cite{bajovic2010distributed, bajovic2012large} where the weight matrices are assumed to be symmetric, stochastic, and random. In contrast, we consider general asymmetric and stochastic weight matrices which are deterministic, and our results imply that optimal error exponent is achieved even if we naively apply the learning rule in~\cite{lalitha2018social} whenever the weight matrix is doubly stochastic. General asymmetric and stochastic weight matrices are also considered in~\cite{bajovic2016distributed}. The main difference from our work is that they focus on optimizing the weight matrix under a given decision region while we achieve the optimal error exponent through modifying the learning rule. 
We provide a decentralized method for estimating the values of the stationary distribution of the consensus Markov Chain. The estimation method only requires bi-directional communication for each pair of nodes forming a directed edge. Meanwhile, optimizing the weight matrix needs to be done globally with a center that knows the entire network topology.

\subsection{Paper Organization}
The rest of this paper is organized as follows. In Section~\ref{sec:formulation}, we formulate our problem and introduce the learning rule proposed in~\cite{lalitha2018social}. In Section~\ref{sec:results}, we propose our modified learning rule and show our main results. The detailed proofs are provided in the appendix. We then propose alternative learning rules for estimating the needed parameters and discuss the convergence of the estimation in Section~\ref{sec:estimating}. In Section~\ref{sec:simulations}, we provide simulation results on the impact of network imbalance, estimation, and quantization. In Section~\ref{sec:discussion}, we further discuss about various aspects of our results including removing the assumptions on the network and extension to the multiple hypothesis testing problems. Finally, we conclude the paper in Section~\ref{sec:conclusion}.

\section{Problem Formulation and Preliminaries} \label{sec:formulation}
\subsection{Problem Formulation}
Consider $n$ nodes collaborating on decentralized binary hypothesis testing. For notational convenience, let $[n]$ denote $\{1,2,...,n\}$. Let $G([n],\mathcal{E})$ denote the underlying directed graph and $\mathcal{N}(i) \triangleq \{ j \in [n]: (i,j) \in \mathcal{E} \}$ denote the neighborhood of node $i$. Node $i$ can get information from node $j$ only if $j \in \mathcal{N}(i)$. To make sure that information can reach all the nodes in the network, we need the following assumption.
\begin{assumption}\label{assu:graph_strongly_connected}
The directed graph $G$ is strongly connected.
\end{assumption}

Regarding the statistical model of the drawn observations at the nodes, let $\mathcal{H}_{\theta}$ denote the hypothesis indexed by $\theta \in [m]$. At each time step $t \in \mathbb{N}$,  each node $i \in [n]$ makes an observation $X_i^{(t)} \in \mathcal{X}_i$, where $\mathcal{X}_i$ denotes the observation space of node $i$ and $\mathcal{X} = \mathcal{X}_1 \times \dots \times \mathcal{X}_n$. Under $\mathcal{H}_{\theta^*}$ being the true hypothesis, the observations $X_i^{(t)} \sim P_{i,\theta^*}$ are i.i.d. over time $t \in \mathbb{N}$, and are independent but not necessarily identical among the nodes. With a slight abuse of notation, we also use $P_{i,\theta}(\cdot)$ to denote the likelihood function for $X_i^{(t)}$ under true hypothesis $\mathcal{H}_\theta$. It should be understood as a probability mass function when the underlying distribution is discrete, and a probability density function when the underlying distribution is absolutely continuous throughout this paper. Each node $i$ is assumed to know its local likelihood functions $P_{i,\theta}(\cdot)$ but not those of other nodes.

\subsection{Social Learning Rule}
\label{sec:social_learning_rule}
In the conventional hypothesis testing problem, the likelihood ratio serves as the optimal statistics in several problems such as the Neyman-Pearson problem and Bayes setting, where the Bayes risk is minimized. The problem in the decentralized case is then whether each node can obtain a statistic that is exactly or close enough to the optimal statistic in the centralized case. A naive approach is that each node simply exchanges its raw observations with others so that each node eventually obtain all the observations among the node. However, the naive approach suffers a high communication cost.

Lalitha \textit{et al.}~\cite{lalitha2018social} proposed a natural approach for decentralized hypothesis testing using the notion of belief propagation. As we will see in later content, the ratio of the beliefs in the proposed learning rule somehow mimics the likelihood ratio but in a slightly tilted form.

Let us describe the proposed learning rule in~\cite{lalitha2018social} as follows. At time step $t$, each node $i\in[n]$ maintains two real vectors: the public belief vector $q^{(t)}_i \in \Delta_m$ and the private belief vector $b^{(t)}_i \in \Delta_m$, which are updated iteratively as $t-1$ changes to $t$. Node $i$ weights the received information from $j$ by $W_{ij}$ which could be seen as the relative confidence that node $i$ has in node $j$.

\begin{enumerate}
    \item Each node $i$ draws an observation $X_i^{(t)} \sim P_{i,\theta^*}$.
    
    \item Each node $i$ updates its public belief vector such that
    \begin{align*}
        b_i^{(t)}(\theta) = \frac{1}{Z^{(t)}_{i,1}} q^{(t-1)}(\theta) P_{i,\theta}(X^{(t)}_i) \quad \forall \theta \in [m],
    \end{align*}
    where $b_i^{(t)}(\theta)$ denotes the $\theta$-th entry of $b_i^{(t)}$.
    
    \item Each node $j$ sends its public belief vector $b_j^{(t)}$ to node $i$ if $j \in \mathcal{N}(i)$.
    
    \item Each node $i$ updates its private belief vector, $q_i^{(t)}$, such that
\[
        q_i^{(t)}(\theta) = \frac{1}{Z^{(t)}_{i,2}} \exp \left\{ \sum_{j=1}^n W_{ij} \log b_j^{(t)}(\theta) \right\} \quad \forall \theta \in [m].
\]
\end{enumerate}
The coefficients $Z^{(t)}_{i,1}, Z^{(t)}_{i,2}$ in steps~2) and~4) normalize the belief vectors such that they fall back into the $m$-dimensional probability simplex.

The results in~\cite{lalitha2018social} show that the entry $q^{(t)}_i(\theta^*)$ converges to one almost surely while the others converge to zero. The rate is also characterized as the weighted sum of the Kullback-Leibler divergence among the distributions over each node.

Though~\cite{lalitha2018social} characterized the convergence performance of the belief vectors, they did not study the probability of error, which seems to be a more concerned perspective in the conventional hypothesis testing problem. As we briefly discussed in Section~\ref{sec:intro}, it can be observed from Figure~\ref{fig:effect_of_imbalance} that if the network tends to be more imbalanced, the learning rule in~\cite{lalitha2018social} tends to suffer a larger gap in the error exponent compared to the centralized case, where a central node is assumed to be capable of gathering all the observations and likelihood functions to perform an optimal test. The gap between the error exponents motivates our modified learning rule introduced in the next section. It turns out that we can close the gap with slight modifications while keeping our learning rule working in a decentralized manner. 

In the following, let us first introduce the log-belief ration test we consider in the rest of our work. 
For the centralized binary detection problem, the randomized likelihood ratio test is optimal (in the Neyman-Pearson problem and the Bayes setting). However, in the decentralized setting, none of the nodes knows the joint likelihood of all the observations in the network and thus no one can carry out the likelihood ratio test. Under the above-mentioned learning rule, we consider the \emph{binary hypothesis testing problem}, and a natural test based on the private belief vector maintained by each node emerges, which is defined as follows.

\begin{definition}[Log-Belief Ratio]
Under the binary hypothesis testing problem, let $\ell^{(t)}_i$ be the (private) log-belief ratio on node $i$ at time $t$ such that
\[
    \ell^{(t)}_i \triangleq \log \frac{q^{(t)}_i(1)}{q^{(t)}_i(0)}.
\]
\end{definition}

\begin{definition}[Log-Belief Ratio Test]
For all $t \in \mathbb{N}$, let $\eta_i^{(t)} \in [0,1]$ and $\gamma_i^{(t)} \in \mathbb{R}$. Define $\varphi_i^{(t)}$ as the log-belief ratio test of node $i$ such that

\begin{align*}
    \varphi_i^{(t)}(\ell_i^{(t)}) \triangleq
    \begin{cases}
        1 &\; \text{if} \; \ell_i^{(t)} > \gamma_i^{(t)}, \\
        \mathrm{Ber}(\eta_i^{(t)}) &\; \text{if} \; \ell_i^{(t)} = \gamma_i^{(t)}, \\
        0 &\; \text{if} \; \ell_i^{(t)} < \gamma_i^{(t)}.
    \end{cases}
\end{align*}
\end{definition}

It is straightforward to see that if there is only a single node, under the learning rule in Section~\ref{sec:social_learning_rule}, the private log-belief ratio $\ell_i^{(t)}$ equals to the log-likelihood ratio, and hence the test is equivalent to the likelihood ratio test.

\section{Main Results}\label{sec:results}
\subsection{Modified Learning Rule}\label{sec:modified}
Our modified learning rule is introduced as follows. At time step $t$, each node $i\in[n]$ maintains two real numbers: the public log-belief ratio $\mu_i^{(t)}$ and the private log-belief ratio $\ell_i^{(t)}$, which are updated iteratively as $t-1$ changes to $t$. Node $i$ weights the received information from $j$ by $W_{ij}$ which could be seen as the relative confidence that node $i$ has in node $j$.

Assume that each node $i \in [n]$ starts with $\ell_{i}^{(0)} = 0$. Let $\theta^*$ denotes the true hypothesis. At each time step $t \in \mathbb{N}$, each node acts as follows:
\begin{enumerate}
    \item Each node $i$ draws an observation $X_i^{(t)} \sim P_{i,\theta^*}$.
    
    \item Each node $i$ updates its public log-belief ratio such that
    \begin{align*}
        \mu_i^{(t)} = \ell_i^{(t-1)} + r_i \log \frac{P_{i,1}(X_i^{(t)})}{P_{i,0}(X_i^{(t)})}
    \end{align*}
    with some $r_i > 0$.
    
    \item Each node $j$ sends its public log-belief ratio $\mu_j^{(t)}$ to node $i$ if $j \in \mathcal{N}(i)$.
    
    \item Each node $i$ updates its private log-belief ratio, $\ell_i^{(t)}$, as the weighted sum of the received $\mu_j^{(t)}$'s such that
    \begin{align*}
        \ell_i^{(t)} = \sum_{j=1}^n W_{ij} \mu_j^{(t)}.
    \end{align*}
\end{enumerate}

\begin{remark}[Equivalence to geometrically weighting the likelihood function]
\label{rem:geometric_weight}
The above learning rule is specialized to binary detection, and it can be extended to general $M$-ary detection problems, the setting originally considered in~\cite{lalitha2018social}. At time $t$, let $q_{i}^{(t-1)}(\theta)$ and $b_i^{(t)}(\theta)$ denote the private and public beliefs for hypothesis $\mathcal{H}_\theta$. The public belief vector then follows the update rule below:
\begin{align*}
    b_{i}^{(t)}(\theta) = \frac{ \left( P_{i,\theta}(X_i^{(t)}) \right)^{r_i} q_{i}^{(t-1)}(\theta) }{ \sum_{\bar{\theta}=1}^M \left( P_{i,\bar{\theta}}(X_i^{(t)}) \right)^{r_i} q_{i}^{(t-1)}(\bar{\theta}) },
\end{align*}
while the private belief vector follows in original update rule in~\cite{lalitha2018social}. Hence, it can be viewed as generalizing the original social learning rule in~\cite{lalitha2018social} by weighting the likelihood function at node $i$ geometrically by $r_i$. Later in Section~\ref{sec:first_order_results} we would see that choosing the weighting vector $r \triangleq (r_1, \dots, r_n)$ properly plays an important role when it comes to optimizing the error exponent. To avoid confusion with the weight matrix $W$, we term $r_i$'s as the \emph{geometric weights} hereafter.
\end{remark}

Now that the additional parameters $r_i$'s emerge, the probability of error depends on the choice of the geometric weights. We formally define them as follows.

\begin{definition}[Probability of Error]
Let $r$ denotes the geometric weights in the learning rule. The type-I and type-II error probabilities for each node $i$ denoted by $\alpha_i^{(t)}(r;\eta_i^{(t)}, \gamma_i^{(t)})$ and $\beta_i^{(t)}(r;\eta_i^{(t)}, \gamma_i^{(t)})$ are defined as
\begin{align}
    \alpha_i^{(t)}(r;\eta_i^{(t)}, \gamma_i^{(t)}) \triangleq&\; \mathsf{Pr} \big\{ \phi_i^{(t)}(\ell_i^{(t)}) = 1 \;\big|\; \mathcal{H}_0 \big\}, \nonumber \\
    \beta_i^{(t)}(r;\eta_i^{(t)}, \gamma_i^{(t)}) \triangleq&\; \mathsf{Pr} \big\{ \phi_i^{(t)}(\ell_i^{(t)}) = 0 \;\big|\; \mathcal{H}_1 \big\}. \nonumber
\end{align}
Note that the performance depends on the chosen geometric weights and the parameters $\eta_i^{(t)}, \gamma_i^{(t)}$.  
\end{definition}

It is then straightforward to come up with a Neyman-Pearson problem for a given choice of the geometric weights:
\begin{alignat*}{2}
    \beta_i^{(t)*}(r,\epsilon) \triangleq \ & \underset{\eta_i^{(t)}, \gamma_i^{(t)}}{\mathsf{minimize}} \;\;&&\; \beta_i^{(t)}(r;\eta_i^{(t)}, \gamma_i^{(t)}) \\
    &\mathsf{subject\,to} &&\; \alpha_i^{(t)}(r;\eta_i^{(t)}, \gamma_i^{(t)}) < \epsilon
\end{alignat*}
for all $i \in [n]$ with some $\epsilon \in (0,1)$. Our goal is to investigate the asymptotic behavior of $\beta_i^{(t)*}(r,\epsilon)$, that is, $\lim_{t \rightarrow \infty} - \frac{1}{t} \log \beta_i^{(t)*}(r,\epsilon). $
For the Bayes risk with prior $(\xi_0,\xi_1)$, we would consider the asymptotic behavior of the Bayes error probability:
\[
\begin{aligned}
    &\mathsf{P}_{\mathsf{e},i}^{(t)*}(r;\xi) \\
    &\triangleq \min_{\eta_i^{(t)}, \gamma_i^{(t)}} \left\{ \xi_0 \alpha_i^{(t)}(r;\eta_i^{(t)}, \gamma_i^{(t)}) + \xi_1 \beta_i^{(t)}(r;\eta_i^{(t)}, \gamma_i^{(t)})  \right\}.
\end{aligned}
\]

\subsection{First-Order Results}\label{sec:first_order_results}
Let us now present our results on the convergence of error probability for the log-belief ratio test under the proposed generalized social learning rule in Section~\ref{sec:social_learning_rule}. 
We begin with error exponents and demonstrate that as long as the geometric weights $r_i$'s are chosen properly, centralized error exponents can be achieved with the proposed decentralized social learning rule. To understand how the price of decentralization kicks in, we further develop results on the higher-order asymptotics and discover that decentralization only costs at most a constant term in the higher-order asymptotics. 

We introduce the results under the assumptions of the corresponding consensus Markov chain being irreducible and aperiodic. Later in Section~\ref{subsec:remove_assump}, we show the necessity of the assumptions and discuss how our learning rule performs when we remove the assumptions.

\begin{assumption}
\label{assum:irreducible_aperiodic}
The $n\times n$ matrix $W$ with the $(i,j)$-th entry being $W_{ij}$ is a transition matrix of some irreducible and aperiodic Markov chain.
\end{assumption}

Let $\pi=[\pi_1 \dots \pi_n]^{\mathsf{T}}$ denote the unique stationary distribution corresponding to the transition matrix $W$. We start with the Neyman-Pearson error exponent for a general choice of the geometric weights.
\begin{theorem}[Neyman-Pearson Error Exponent for General Geometric Weights]
\label{thm:decentralized_optmization_form}
Suppose that Assumptions~\ref{assu:graph_strongly_connected} and~\ref{assum:irreducible_aperiodic} hold. For the Neyman-Pearson problem, the type-II error exponent for each node $i$ is characterized as shown on the top of the next page.
\end{theorem}

\begin{figure*}[!t]
\normalsize
\begin{align*}
\forall\, \epsilon > 0,\ 
\lim_{t \rightarrow \infty} - \frac{1}{t} \log \beta_i^{(t)*}(r,\epsilon)
= \sup_{\lambda \geq 0} \left\{ 
\sum_{j=1}^n \lambda \pi_j r_j \mathsf{E}_{P_0} \left[ \log \frac{P_{j,0}(X_j)}{P_{j,1}(X_j)} \right] - \log \mathsf{E}_{P_1} \left[ \left( \frac{P_{j,0}(X_j)}{P_{j,1}(X_j)} \right)^{\lambda \pi_j r_j} \right] 
\right\},
\ i\in[n].
\end{align*}
\hrulefill
\vspace*{4pt}
\end{figure*}

\begin{sketch}
The error probability is the probability of the statistic, the sum of the log-likelihood ratios, falling into the wrong decision region, where the threshold could be proved to keep the type-I error under the constant constraint with the weak law of large numbers. Since the log-likelihood ratios are mutually independent over time and across nodes, the error exponent is characterized by the large deviation theorems (Gärtner-Ellis Theorem). Through simplifying the optimization term in the large deviation theorem, our theorem is proved. The detailed proof is provided in Appendix~\ref{pf:decentralized_optmization_form}.
\end{sketch}

Theorem~\ref{thm:decentralized_optmization_form} shows the error exponent for any choice of the geometric weights. If we have $r_i = 1$ for all $i \in [n]$, Theorem~\ref{thm:decentralized_optmization_form} gives the error exponent using the original learning rule proposed in~\cite{lalitha2018social}. It turns out that the optimal error exponent for the centralized case could be achieved through a proper choice of $r$, which is shown in the following theorem. This suggests that social learning is as good as centralized detection in terms of the error exponent in the Neyman-Pearson problem.

\begin{theorem}[Social Learning is Almost as Good as Centralized Detection in Neyman-Pearson Problem]
\label{thm:neyman_pearson_exponent}
Suppose that Assumptions~\ref{assu:graph_strongly_connected} and~\ref{assum:irreducible_aperiodic} hold. If each agent $i$ knows $\pi_i$, the value of the stationary distribution of the weight matrix at that node, by choosing $r = r^*$ where $r^*_i = c/\pi_i$ for all $i$ and some common constant $c \in \mathbb{R}$ among the nodes, we have
\begin{align*}
\lim_{t \rightarrow \infty} - \frac{1}{t} \log \beta_i^{(t)*}(r^*, \epsilon)
&= \sum_{j=1}^n D_{\mathrm{KL}} \left( P_{j,0} \Vert P_{j,1} \right)\\
&= D_{\mathrm{KL}} \left( P_{0} \Vert P_{1} \right)\quad \forall\, i \in [n].
\end{align*}
Here $P_\theta$ denotes the product distribution of $P_{1,\theta}, ..., P_{n,\theta}$ and $D_{\mathrm{KL}}$ denotes the Kullback–Leibler divergence.
\end{theorem}

\begin{sketch}
By plugging $r = r^*$ into Theorem~\ref{thm:decentralized_optmization_form}, we identify the optimization term as a variational representation of the Kullback-Leibler divergence. With $\lambda^*=1$ being the maximizer of the optimization problem, we can show that choosing the appropriate geometric weights leads us to the same error exponent we see in the centralized case. The detailed proof can be found in Appendix~\ref{pf:neyman_pearson_exponent}.
\end{sketch}

The original social learning rule suffers from the unfairness in consensus. Since we are focusing on the asymptotic result for a given graph, the information on each node must have sufficient time to propagate to any other nodes in the graph. All we need to do is to carefully re-weight the log-likelihood ratios such that each observation is equally important, which allows the network to attain the optimal error exponent.

The optimal weight is proportional to the inverse of the local stationary distribution. It means that if node $i$ is not trusted by the other nodes such that $\pi_i$ is relatively small, then node $i$ should amplify its messages to make its observations as important as anyone else's. Hence, node $i$ should weigh the log-likelihood ratio with $r_i = \pi_i^{-1}$ before infusing it into the network to equalize the gain due to the unfair consensus. If the stationary distribution is uniform, which is the case when the weight matrix is doubly stochastic, each node's observations are equally important by nature and thus the optimal error exponent is obtained by simply applying the learning rule in~\cite{lalitha2018social}.

Furthermore, the theorem suggests that even if some of the nodes have larger Kullback–Leibler divergence terms, which means that they have better capabilities of distinguishing the hypotheses, their information should not be more important than anyone else's.

For the Bayes risk, by choosing the geometric weight $r=r^*$, the same choice as in Theorem~\ref{thm:neyman_pearson_exponent}, we also show that the centralized error exponent is attained.

\begin{theorem}[Social Learning is Almost as Good as Centralized Detection under the Bayes Setting]
\label{thm:bayes_exponent}
Suppose that Assumptions~\ref{assu:graph_strongly_connected} and~\ref{assum:irreducible_aperiodic} hold. If each agent $i$ knows $\pi_i$, the value of the stationary distribution of the weight matrix at that node, by choosing $r = r^*$, for all prior $\xi$, we have
\begin{align*}
\lim_{t \rightarrow \infty} - \frac{1}{t} \log \mathsf{P}_{\mathsf{e},i}^{(t)*}(r^*, \xi) = \mathrm{CI}(P_0, P_1),
\end{align*}
where $\mathrm{CI}(P_0,P_1)$ denotes the Chernoff information between $P_0$ and $P_1$, that is, 
\begin{align*}
&\mathrm{CI}(P_0, P_1)\\
&= \max_{\lambda \in [0,1]} \left\{- \log \mathsf{E}_{P_0} \left[ \left( \frac{P_1(\pmb{X})}{P_0(\pmb{X})} \right)^{\lambda} \right] \right\}\\
&= \max_{\lambda \in [0,1]} \left\{- \log \mathsf{E}_{P_0} \left[ \left( \frac{P_{1,1}(X_1) \dots P_{n,1}(X_n)}{P_{1,0}(X_1) \dots P_{n,0}(X_n)} \right)^{\lambda} \right] \right\}.
\end{align*}
\end{theorem}

\begin{sketch}
We prove this with a similar technique we used in the proof of Theorem~\ref{thm:decentralized_optmization_form}, but now we simply set the threshold for the testing to zero. Details are provided in Appendix~\ref{pf:bayes_exponent}.
\end{sketch}

\subsection{Higher-Order Asymptotics}\label{sec:higher-order}
While Theorem~\ref{thm:neyman_pearson_exponent} shows that decentralization does not affect the error exponent if we choose the geometric weights properly, further investigation on the probability of error may reveal the effect of decentralization. 
To characterize the higher-order asymptotics, we further impose the following assumption on the log-likelihood ratios.

\begin{assumption}[Bounded Log-Likelihood Ratios]
\label{assum:bounded_llr}
The log-likelihood ratio is bounded by some constants $L_1, \dots, L_n > 0$ for each node $i$, that is,
\begin{align*}
\left\vert \log \frac{ P_{i,1}(x_i) }{ P_{i,0}(x_i) } \right\vert \leq L_i \quad \forall x_i \in \mathcal{X}_i \; \forall i \in [n].
\end{align*}
\end{assumption}

The assumption holds whenever the support $\mathcal{X}_i$ is finite for each node $i \in [n]$. Another assumption on the network is also made as follows.

\begin{assumption}
\label{assum:reversible}
The consensus Markov chain is reversible.
\end{assumption}

Under Assumption~\ref{assum:reversible}, the convergence of the power of $W$ is captured by $\rho \triangleq\max \{ \lambda_2, \lvert \lambda_n \rvert \}$, where $\lambda_k$ is the $k$-th largest eigenvalue of $W$, $k\in[n]$. Since the spectral gap, $1-\rho$, is related to the connectivity of the network, it is intuitive that the term emerges in the price of decentralization. 

The following theorem reveals the effect of decentralization.
\begin{theorem}[The Effect of Decentralization in Neyman-Pearson Problem] \label{thm:neyman_pearson_higher_order}
Suppose that Assumptions~\ref{assu:graph_strongly_connected}, \ref{assum:irreducible_aperiodic}, \ref{assum:bounded_llr}, and \ref{assum:reversible} hold. Assume that each distribution of the log likelihood ratio, $\log \frac{ P_{i,1}(X_i) }{ P_{i,0}(X_i) }$, is non-lattice for $i \in [n]$. By choosing $r=r^*$, the type-II error probability is upper bounded as
\begin{align*}
\beta^{(t)*}_{\mathrm{cen}}(\epsilon) \leq \beta^{(t)*}_i(r^*, \epsilon) \leq C^{(\mathrm{NP})}_i \,\beta^{(t)*}_{\mathrm{cen}}(\epsilon),
\end{align*}
where $\beta^{(t)*}_{\mathrm{cen}}(\epsilon)$ is the optimal type-II error probability in the centralized case and the expression of the constant $C^{(\mathrm{NP})}_i$ is shown on the top of the next page. 
\end{theorem}

\begin{sketch}
We evaluate the higher-order asymptotics in the exponent of the error probability. Through the change of measure among the two distributions under the two hypotheses, the distribution of the sum of the log-likelihood ratio is distributed around the threshold for testing, and this is where Esseen's theorem in~\cite{esseen1945fourier} gives us a more detailed analysis. Together with convergence results on Markov chains, our theorem is proved. The full proof is provided in Appendix~\ref{pf:neyman_pearson_higher_order}.
\end{sketch}

\begin{figure*}[!t]
\normalsize
\begin{align}\label{eq:const_pen}
C^{(\mathrm{NP})}_i = \exp \left\{ \frac{\rho}{1-\rho} \sqrt{\frac{1-\pi_i}{\pi_i}} \left( \sqrt{\sum_{j=1}^n \frac{1}{\pi_j} \left( D_{\mathrm{KL}} \left( P_{j,0} \Vert P_{j,1} \right) \right)^2 } + \sqrt{\sum_{j=1}^n \frac{1}{\pi_j} L_j^2 } \right) + o(1) \right\},\quad i\in[n].
\end{align}
\hrulefill
\vspace*{4pt}
\end{figure*}

For the constant penalty in Theorem~\ref{thm:neyman_pearson_higher_order}, the term $\rho$ in \eqref{eq:const_pen} represents the connectivity of a graph such that a graph with higher connectivity (smaller $\rho$) obtains a smaller upper bound on the probability of error. For example, a complete graph with self-loops obtains the weight matrix $W = \frac{1}{n} \pmb{1} \pmb{1}^{\mathsf{T}}$ by making each node uniformly distributing weights to its neighbors. In this case, we have $\rho=0$ and thus the network obtains no constant penalty. Meanwhile, for a ring with each node giving both its neighbor half of its confidence, we have $\rho=\cos\frac{2\pi}{n}$ and the network suffers a larger price of decentralization as the network size $n$ grows.

\begin{figure}[!t]
\centering
\includegraphics[width=.8\linewidth]{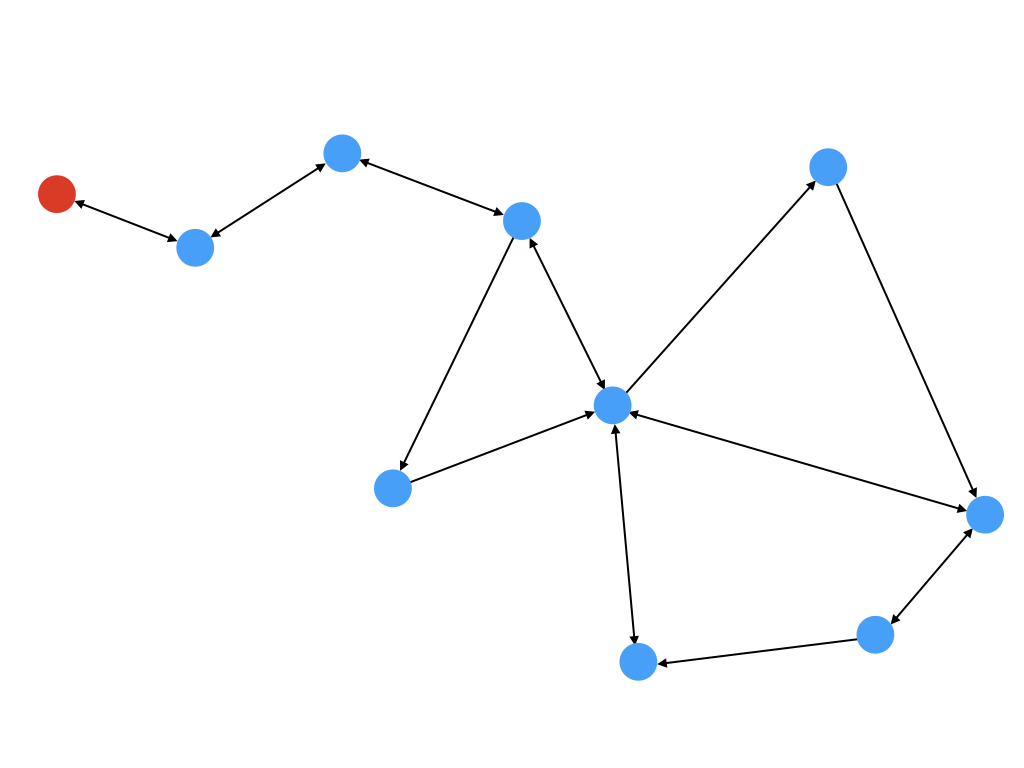}
\caption{A node (colored red) with small $\pi_i$.}
\label{fig:isolated_node}
\end{figure}

Notice that the constant penalty may differ among the nodes with the stationary distribution at each node. Generally speaking, a node with a smaller corresponding $\pi_i$ gathers information slower since it gains less trust from the network compared to the others. For example, in Figure~\ref{fig:isolated_node}, a node that is far from the others in the network tends to obtain s smaller stationary distribution at it and suffer a larger price of decentralization.

The reason why the price of decentralization emerges as a constant term in the exponent of the error probability is also quite intuitive. Since we focus on the asymptotic analysis with respect to $t$ while the size of the network remains fixed, information needs only constant time to travel through the network regardless of the value of $t$. In fact, we can re-write Theorem~\ref{thm:neyman_pearson_higher_order} into the following form.

\begin{corollary}[Viewing the Price of Decentralization as a Constant Time Delay for Decentralized Testing]
\label{cor:time_delay}
It is straightforward to see that from Theorem~\ref{thm:neyman_pearson_higher_order} we have
\begin{align*}
    \beta^{(t)*}_{\mathrm{cen}}(\epsilon) \leq \beta^{(t)*}_i(r^*, \epsilon) \lesssim \beta^{(t-d_i)*}_{\mathrm{cen}}(\epsilon),
\end{align*}
where
\begin{align*}
    d_i = \frac{\log C^{(\mathrm{NP})}_i}{\sum_{j=1}^n D_{\mathrm{KL}} \left( P_{j,0} \Vert P_{j,1} \right)},\ i\in [n], 
\end{align*}
is a constant with respect to $t$. The notation $a^{(t)} \lesssim b^{(t)}$ means that $\lim_{t \rightarrow \infty} \frac{a^{(t)}}{b^{(t)}} \leq 1$.
\end{corollary}

\begin{proof}
The result follows from the proof of Theorem~\ref{thm:neyman_pearson_higher_order} in Appendix~\ref{pf:neyman_pearson_higher_order}.
\end{proof}

Corollary~\ref{cor:time_delay} provides another perspective toward the price of decentralization. While each node surely could not outperform the centralized case, it outperforms the centralized case with additional $d_i$ observations and rounds of communications. The additional number of rounds being a constant with respect to $t$ follows the similar intuition mentioned above before Corollary~\ref{cor:time_delay}.

A similar result is obtained for the Bayes risk.

\begin{theorem}[The Effect of Decentralization on the Bayes risk]
\label{thm:bayes_higher_order}
Suppose that Assumptions~\ref{assu:graph_strongly_connected}, \ref{assum:irreducible_aperiodic}, \ref{assum:bounded_llr}, and \ref{assum:reversible} hold. 
Assume that each distribution of the log likelihood ratio, $\log \frac{ P_{i,1}(X_i) }{ P_{i,0}(X_i) }$, is non-lattice for $i \in [n]$. By choosing $r=r^*$, the Bayes risk is bounded as
\begin{align*}
    \mathsf{P}^{(t)*}_{\mathsf{e}, \mathrm{cen}}(\xi) \leq \mathsf{P}^{(t)*}_{\mathsf{e}, i}(r; \xi) \lesssim C^{(\mathrm{B})}_i \, \mathsf{P}^{(t)*}_{\mathsf{e}, \mathrm{cen}}(\xi)
\end{align*}
where $\mathsf{P}^{(t)*}_{\mathsf{e}, \mathrm{cen}}(\xi)$ is the optimal Bayes risk in the centralized case and the expression of the constant $C^{(\mathrm{B})}_i$, $i\in [n]$, is shown below:
\[
\textstyle
\exp \left\{ \max(\theta^*,1-\theta^*)\frac{\rho}{1-\rho}\sqrt{\frac{1-\pi_i}{\pi_i} \left( \sum_{j=1}^n \frac{1}{\pi_j} L_j^2 \right) }+ o(1) \right\}, 
\]
with $\theta^* = \underset{\theta \in [0,1]}{\arg \max}\ \left\{- \log \mathsf{E}_{X \sim P_0} \left[ \left( \frac{P_1(X)}{P_0(X)} \right)^{\theta} \right]\right\}$.
\end{theorem}
\begin{sketch}
The proof is similar to the one of Theorem~\ref{thm:neyman_pearson_higher_order}. However, in the Bayes case, we change the measure not among the two distributions of the hypotheses, but to the exponentially tilted distribution of the two distributions. In this case, the distribution of the sum of the log-likelihood ratios is located around the threshold, which is zero. The detailed proof is provided in Appendix~\ref{pf:bayes_higher_order}.
\end{sketch}

In Theorem~\ref{thm:neyman_pearson_higher_order} and Theorem~\ref{thm:bayes_higher_order}, the upper bounds hint that the effect of decentralization depends on the network connectivity with the term $\frac{\rho}{1-\rho}$. In the special case that $X_i^{(t)}$ follows Gaussian distributions for all $i \in [n]$, the optimal type-II error probability in the Neyman-Pearson problem is characterized, and it shows that the effect of decentralization relates to the network connectivity with the term $\frac{\rho^2}{1-\rho^2}$. When $\rho$ goes to zero, the term $\frac{\rho^2}{1-\rho^2}$ goes to zero faster than $\frac{\rho}{1-\rho}$, that is, in the Gaussian case, $\frac{\rho^2}{1-\rho^2}$ is a tighter characterization of the effect of decentralization. However, notice that in this case, our Assumption~\ref{assum:bounded_llr} is not satisfied. The result hints that our characterization of the effect of decentralization is either incomplete due to the assumption or not tight enough.

\begin{theorem}[The Effect of Decentralization in Neyman-Pearson Problem for the Special case of Gaussian distributions]
\label{thm:gaussian}
Suppose that Assumptions~\ref{assu:graph_strongly_connected}, \ref{assum:irreducible_aperiodic}, and \ref{assum:reversible} hold. 
Assume that for each node $i$, the observations follow the i.i.d. Gaussian distributions, that is,
\begin{align*}
\mathcal{H}_0 : X^{(t)}_i &\overset{\text{i.i.d.}}{\sim} \mathrm{Normal}(-\mu, \sigma^2) \\
\mathcal{H}_1 : X^{(t)}_i &\overset{\text{i.i.d.}}{\sim} \mathrm{Normal}(\mu, \sigma^2).
\end{align*}
Then, in the Neyman-Pearson problem, the optimal type-II error is upper bounded as
\begin{align*}
    \beta^{(t)*}_i(r^*,\epsilon)
    \leq C^{(\mathrm{B,Gaussian})}_i \, \beta^{(t)*}_{\mathrm{cen}}(\epsilon)
\end{align*}
where
\begin{align*}
    C^{(\mathrm{B,Gaussian})}_i = \frac{2\mu}{\sigma} \left( \frac{\rho^2}{1-\rho^2} \right) \left( \frac{\pi_i}{1-\pi_i} \right) \left( \sum_{j=1}^n \frac{1}{\pi_j} \right).
\end{align*}
\end{theorem}

\begin{sketch}
For Gaussian observations, the log-likelihood ratios follow another Gaussian distribution as well, and thus we characterize the optimal threshold with the inverse Q-function and plug the threshold in to directly calculate the corresponding type-II error probability. The rest are to approximate the Q-function and control the deviation of the Markov chain with some upper bounds. The detailed proof is provided in Appendix~\ref{pf:NP_gaussian}.
\end{sketch}

\section{Obtaining the Geometric Weights}\label{sec:estimating}
\subsection{Estimating the Optimal Geometric Weights in a Decentralized Way}\label{sec:estimation}

Theorem~\ref{thm:neyman_pearson_exponent} and Theorem~\ref{thm:bayes_exponent} state that the optimal error exponent can be attained at each node as long as each node~$i$ has access to $\pi_i$, the $i$-th entry of the stationary distribution of the Markov chain whose transition matrix is the weight matrix $W$. For each node to learn the local stationary distribution, some additional effort is needed in the decentralized setting. The most naive way is to request a center that knows the entire weight matrix $W$ to calculate the stationary distribution and disseminate the corresponding information to each node. However, such a centralized method is not desirable from the perspective of decentralization. 

Let us provide a simple iterative and decentralized estimation algorithm. Recall that in our problem formulation, node~$i$ grabs information from node~$j$ only if $j \in \mathcal{N}(i)$. For the estimation, we make an additional yet practical assumption that node~$i$ is able to send information to node~$j$ if $j \in \mathcal{N}(i)$. In other words, communication takes place bidirectionally. 

The algorithm is described as follows. Let each node~$i$ maintain a real numbers $\hat{\pi}^{(t)}_i$ and randomly initialize the value such that $\hat{\pi}^{(0)}_i>0$. At round~$t=1,2,\dots$, each node~$i$ multiplies $\hat{\pi}^{(t-1)}_i$ by $W_{ij}$ to form the message $\nu^{(t)}_{ij}$ and send it to node~$j$ for further consensus, that is, $\hat{\pi}^{(t)}_j = \sum_{i: (i,j)\in\mathcal{E}} \nu^{(t)}_{ij} = \sum_{i: (i,j)\in\mathcal{E}} W_{ij} \hat{\pi}^{(t-1)}_i$.

It turns out that if $W$ corresponds to the transition matrix of a reversible Markov chain, the local estimation converges to $s \pi_i$ exponentially fast with (recall $\rho$ is defined in Section~\ref{sec:first_order_results}) with $s=\sum_{i=1}^n \hat{\pi}_i^{(0)}$ being the sum of the initial values. That is,
\begin{align*}
    \sum_{i=1}^n \left\vert \hat{\pi}^{(T)}_i - s \pi_i \right\vert
    \leq \left( \sum_{i=1}^n \sqrt{\frac{1-\pi_i}{\pi_i}} \right) \left( \sum_{i=1}^n \pi_i \left( \hat{\pi}_i^{(0)} \right)^2 \right) \rho^{T}.
\end{align*}
This is proved by the following argument: with Lemma~\ref{lem:markov_convergence} in Appendix~\ref{pf:neyman_pearson_higher_order}, we have
\begin{align*}
    \sum_{j=1}^n \left\vert \hat{\pi}^{(T)}_j - s \pi_j \right\vert
    &= \sum_{j=1}^n \left\vert \left( \sum_{i=1}^n \hat{\pi}_i^{(0)} [W^T]_{ij} \right) - s \pi_i \right\vert \\
    &\leq \sum_{i=1}^n \sum_{j=1}^n \hat{\pi}_i^{(0)} \left\vert [W^T]_{ij} - \pi_i \right\vert\\
    &\leq \left( \sum_{i=1}^n \sqrt{\frac{1-\pi_i}{\pi_i}} \right) \left( \sum_{i=1}^n \pi_i \left( \hat{\pi}_i^{(0)} \right)^2 \right) \rho^{T}.
\end{align*}

Notice that the factor $s$ does not matter since each node is not required to know the exact value of the corresponding entry in the stationary distribution. Instead, as we showed in Theorem~\ref{thm:neyman_pearson_higher_order} and Theorem~\ref{thm:bayes_higher_order}, any common constant among the choices of the geometric weights on each node is innocuous to our results.

\begin{figure}[!ht]
     \centering
     \includegraphics[width=\linewidth]{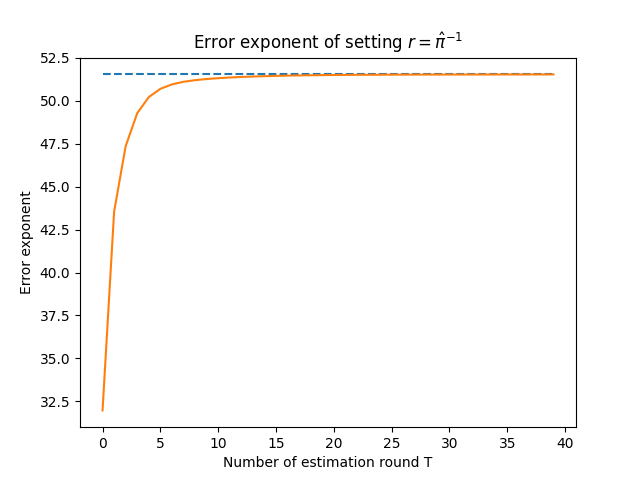}
     \caption{Convergence of the error exponent.}
     \label{fig:estimation}
\end{figure}

\begin{figure}[!ht]
     \centering
     \includegraphics[width=\linewidth]{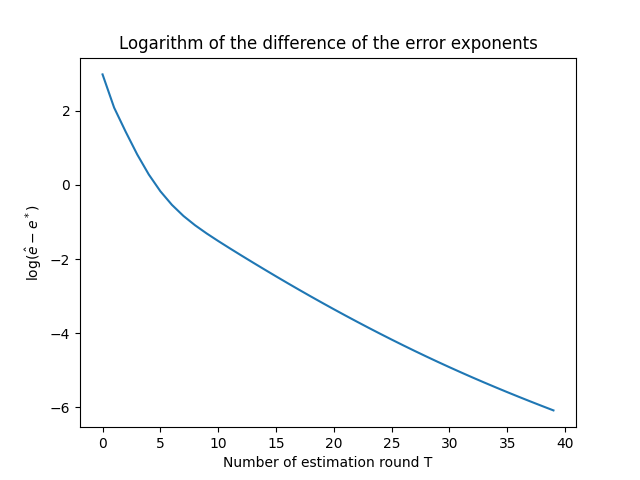}
     \caption{Convergence of the error exponent in log scale.}        
     \label{fig:estimation_log}
\end{figure}

After estimating the stationary distribution for $T$ rounds, the nodes then adopt the social learning rule with geometric weight $r_i = \left( \hat{\pi}^{(T)}_i \right)^{-1}$ for all $i \in [n]$. To illustrate the achievable error exponent of this ``plug-in'' social learning rule, let
\begin{align*}
    \hat{\mathrm{e}}_T = \lim_{t \rightarrow \infty} - \frac{1}{t} \log \beta_i^{(t)*} \left( \left( \hat{\pi}^T \right)^{-1}, \epsilon \right).
\end{align*}
In Figure~\ref{fig:estimation} and Figure~\ref{fig:estimation_log}, we illustrate the convergence of $\hat{\mathrm{e}}_T$ to the centralized error exponent through a random scale-free network with 100 nodes and Bernoulli sources $X_{i}^{(t)} \sim \mathrm{Ber}(q_{i,\theta})$, $i=1,2$, $\theta = 0,1$. The numerical result hints that $\hat{\mathrm{e}}_T$ converges to the optimal error exponent exponentially fast with $T$.

\subsection{Combining the Learning Rule with the Estimation of the Optimal Geometric Weights}
\label{sec:alter}
In Section~\ref{sec:estimation}, we provide a simple method for estimating the stationary distribution in a decentralized manner. Though the estimation error vanishes exponentially, a first-order loss remains due to the difference between the estimation and the true stationary distribution. However, we can keep estimating the stationary distribution while executing the social learning rule, and the combined learning rule becomes the following.
\begin{itemize}
    \item First, estimate the stationary distribution for $T_{\mathsf{E}}$ rounds. Let $\hat{\pi}^{(0)}_i = 0$ for all $i \in [n]$ and for $t=1, 2, \dots, T_{\mathsf{E}}$, each node $i$ does the following steps.
    \begin{enumerate}
        \item Send the message $W_{ij} \hat{\pi}^{(t-1)}_i$ to each node $j \in \mathcal{N}(i)$.
        \item Update the estimation with the received messages such that
        \begin{align*}
            \hat{\pi}^{(t)}_i = \sum_{j:i \in \mathcal{N}(j)} W_{ji} \hat{\pi}^{(t-1)}_j.
        \end{align*}
    \end{enumerate}
    \item Then, keep estimating while executing our learning rule for $T_{\mathsf{EL}}$ rounds. Let $\ell^{(0)}_i=0$ for all $i \in [n]$ and for $t=1, 2, \dots, T_{\mathsf{EL}}$, each node $i$ does the following steps.
    \begin{enumerate}
        \item Each node $i$ draws an observation $X^{(t)}_i \sim P_{i, \theta^*}$ 
        \item Each node $i$ updates its public log-belief ratio such that
        \begin{align*}
            \mu^{(t)}_i = \ell^{(t-1)}_i + \left( \hat{\pi}^{(T_{\mathsf{E}}+t-1)}_i \right)^{-1} \log \frac{P_{i,1}(X^{(t)}_i)}{P_{i,0}(X^{(t)}_i)}
        \end{align*}
        \item For each $j \in \mathcal{N}(i)$, node~$i$ sends $W_{ij} \hat{\pi}^{(T_\mathsf{E}+t-1)}_i$ to node~$j$ and get $\mu^{(t)}_j$ from node~$j$.
        \item Each node~$i$ updates its estimation and private log-belief ratio.
        \begin{align*}
            \hat{\pi}^{(T_\mathsf{E}+t)} &= \sum_{j:i \in \mathcal{N}(j)} W_{ji} \hat{\pi}^{(T_\mathsf{E}+t-1)}_j, \\
             \ell_i^{(t)} &= \sum_{j=1}^n W_{ij} \mu_j^{(t)}.
        \end{align*}
    \end{enumerate}
\end{itemize}

In the first part, we estimate the stationary distribution for $T_{\mathsf{E}}$ rounds, and in the second part, we adopt our learning rule together with estimating the stationary distribution iteratively. We conjecture that the combined learning rule achieves the optimal error exponent, which might be shown with a proof similar to the one of Theorem~\ref{thm:neyman_pearson_exponent}. In the next section, the performance of the proposed scheme is demonstrated through simulations.

\section{Simulations} \label{sec:simulations}
\subsection{Impact of Network Imbalance}
We have shown that how network imbalance can impact the error exponent through Theorem~\ref{thm:decentralized_optmization_form} and Figure~\ref{fig:effect_of_imbalance}. In the following we provide a simulation for the impact of network imbalance to support our statement regarding the non-asymptotic performance.

\begin{figure*}[!t]
    \normalsize
    \centering
    \includegraphics[width=\textwidth]{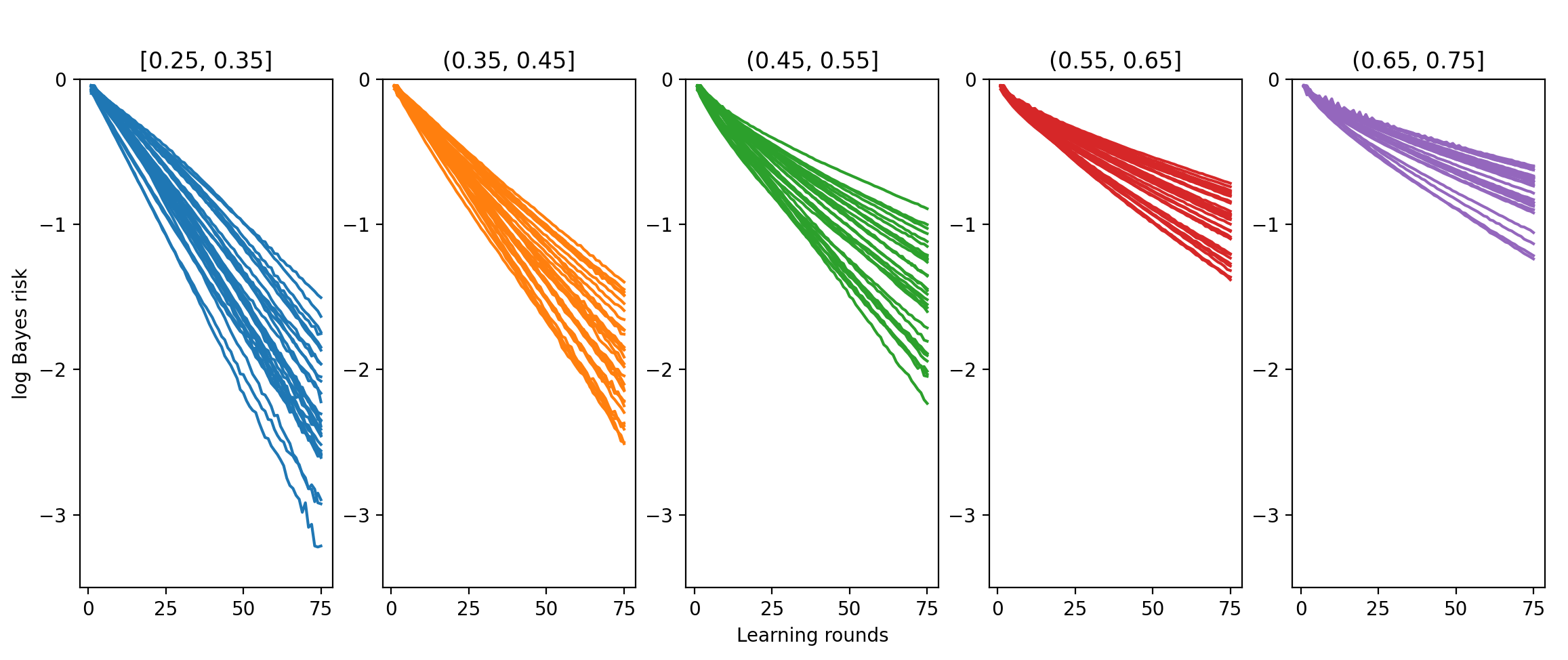}
    \caption{Impact of network imbalance.}
    \label{fig:imbalance}
\end{figure*}

In Figure~\ref{fig:imbalance}, random scale-free networks with 30 nodes in each are sampled. Each node in the sampled network distributes its relative confidence uniformly to its neighbors to form the weight matrix, and the corresponding stationary distribution is evaluated. Each sampled network is tagged with its quantity of imbalance, which is the total variation between its corresponding stationary distribution and the uniform one, and clustered into five groups. The simulation results for these groups of networks are gathered in the five subplots accordingly. For example, the leftmost subplot gathers the results of 30 sampled networks with quantities of imbalance falling within 0.25 to 0.35. Notice that we use the total variation between the stationary distribution and the uniform one to quantify the network imbalance, however, one can use other distance measurements such as the vector 2-norm as in Figure~\ref{fig:effect_of_imbalance}.

Each time a network is sampled, random observations are drawn on each node and follow the Bernoulli distributions with parameters 0.5 or 0.6 depending on the underlying true hypothesis. We utilize the learning rule in Section~\ref{sec:alter} and record the testing result at each iteration. Such procedure repeats 100,000 times for each network and we get an empirical result on Bayes risk over time.

Figure~\ref{fig:imbalance} shows that for networks with lower quantities of imbalance, learning is more efficient. The Bayes risks vanish a lot faster than the ones for networks with higher quantities of imbalance. Although we did not evaluate an explicit relation between the network imbalance and the probability of error, Figure~\ref{fig:imbalance} does roughly show us the trade-off between them.

\subsection{Compensating the Network Imbalance}
Our main theorems show that our learning rule does obtain the optimal error exponent. For the non-asymptotic performance, we show that our learning rule obtains great improvement on the probability of error by compensating the network imbalance compared to the original learning rule in~\cite{lalitha2018social}.

\begin{figure}[!t]
    \centering
    \includegraphics[width=\linewidth]{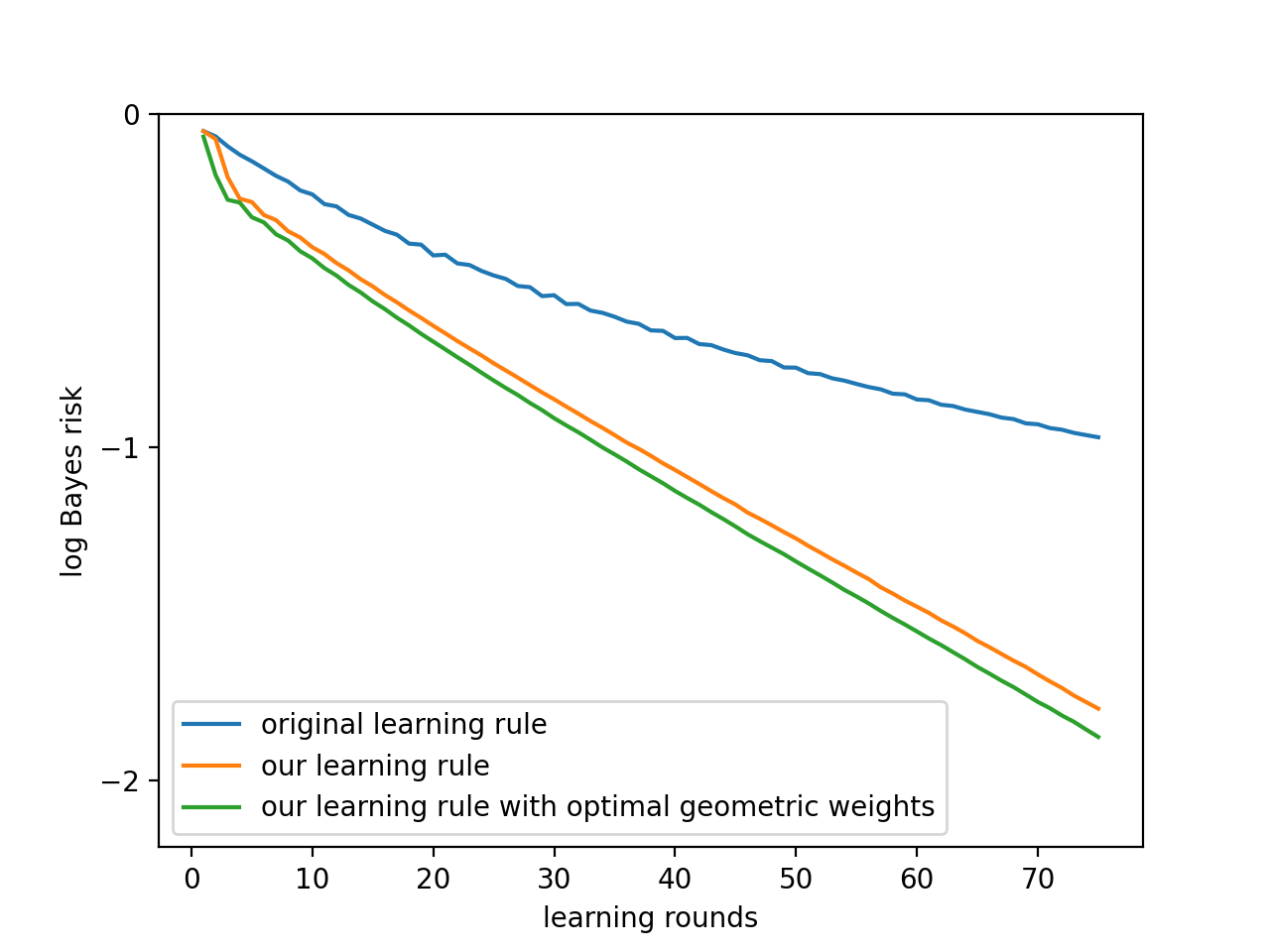}
    \caption{Comparing the learning rules.}
    \label{fig:compensate}
\end{figure}

We sampled 1,000 random scale-free networks with 50 nodes in each. Each node draws random samples and we record the error events under:
\begin{enumerate}
    \item The learning rule in~\cite{lalitha2018social}.
    \item Our learning rule in Section~\ref{sec:alter}.
    \item Our learning rule with each node knowing the stationary distribution \textit{a priori}.
\end{enumerate}
We set the learning rounds to 75 and repeat the procedure 1,000 times to get the empirical Bayes risks. We compute the log Bayes risks for each random network and show the average log Bayes risk over the networks in Figure~\ref{fig:compensate}.

Figure~\ref{fig:compensate} shows that the original learning rule in~\cite{lalitha2018social} suffers a slower decrease in the probability of error. The green line indicates the case that each node $i$ knows $\pi$ in prior and sets its geometric weight to be $r_i = \pi_i^{-1}$. The yellow line represents our learning rule in Section~\ref{sec:alter} in which each node keeps estimating the stationary distribution for its choice of geometric weight while executing our learning rule in Section~\ref{sec:modified}. Our result (yellow line) has a much more rapid decrease in the probability of error compared to the one with the original learning rule in~\cite{lalitha2018social}. We can see that the yellow line takes a few rounds to divert and move along with the green line since the estimation on the stationary distribution for each node soon converges close enough to the true stationary distribution multiplied by $n$.

The result shows that our learning rule not only guarantees the optimal error exponent, but it performs well in non-asymptotic scenarios.

\subsection{The Effect of Initial and Ongoing Estimations}\label{subsec:sim_est_soc_lrn}
In the method proposed in Section~\ref{sec:alter}, we estimate the stationary distribution for several rounds before each node starts drawing observations. Furthermore, we can choose whether to keep estimating the stationary distribution. In the following, we demonstrate the effect of such initial and ongoing estimations on the stationary distribution.

\begin{figure}[!t]
    \centering
    \includegraphics[width=\linewidth]{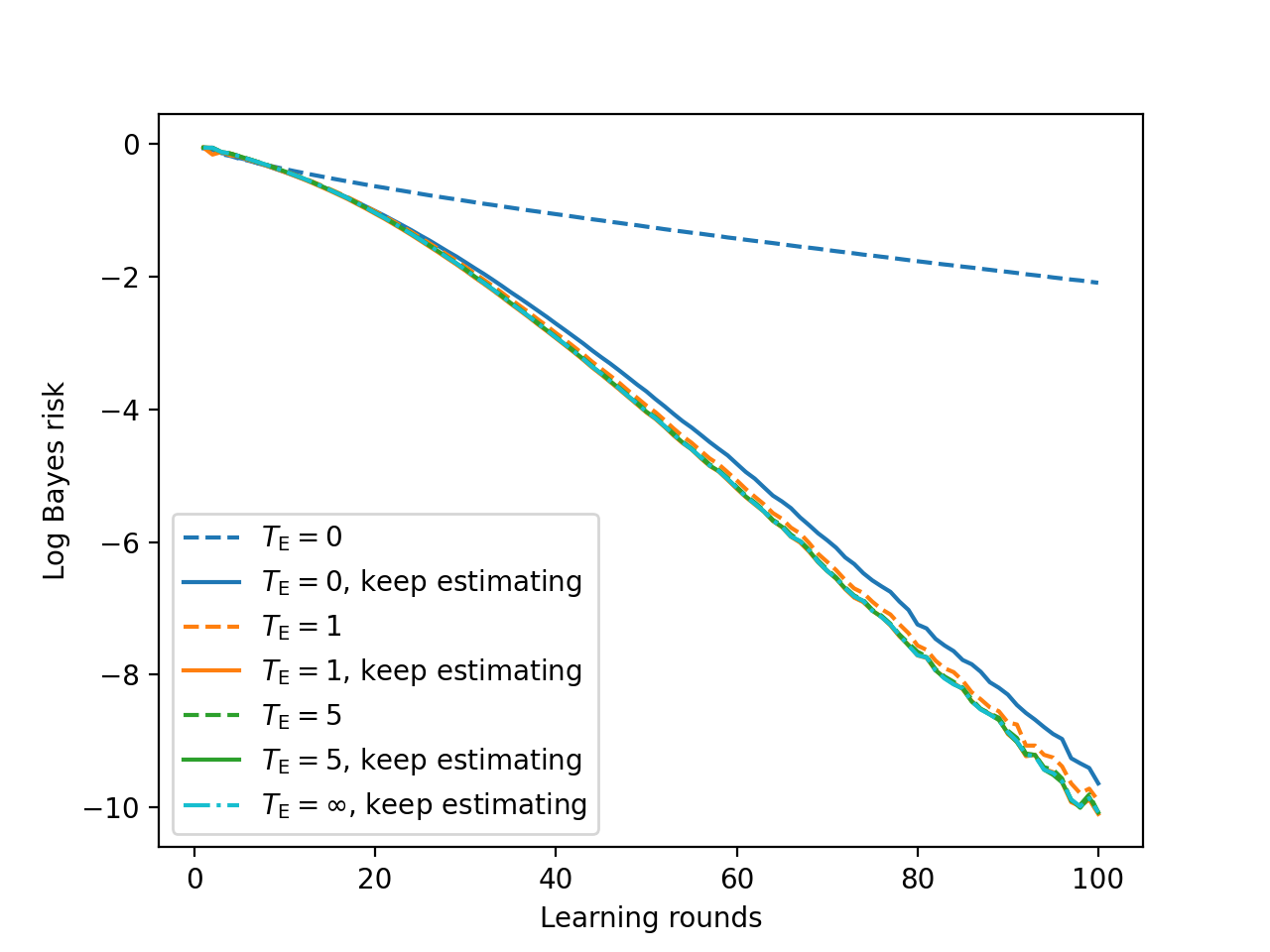}
    \caption{The effect of initial and ongoing estimations.}
    \label{fig:estLrn}
\end{figure}

In Figure~\ref{fig:estLrn}, simulation results of two random scale-free networks with 100 nodes in each are shown. The quantity $T_\mathsf{E}$ is defined in Section~\ref{sec:alter} to be the number of initial estimation rounds. The dashed lines represent the log Bayes risks under the cases without keeping estimating the stationary distribution after each node starts drawing the observations, and the solid lines correspond to the cases where each node keeps estimating the stationary distribution for its choice of the geometric weight. As we can see in the figure, the blue dashed line represents the Bayes risk with neither initial nor ongoing estimations, which is equivalent to applying the original learning rule in~\cite{lalitha2018social} where the geometric weights are set to be one uniformly. The blue solid line also has $T_\mathsf{E}$ to be zero, while in this case, each node keeps estimating the stationary distribution to form its choice of the geometric weight. We can see that the blue solid line obtains a great improvement in terms of the log Bayes risk compared to the blue dashed line.

The light blue line represents the result for $T_\mathsf{E}=\infty$, which means that each node knows the exact stationary distribution in prior. We can see that other results with $T_\mathsf{E}>0$ perform closely to the light blue line, except for the orange dashed line showing a relatively minor gap with them.

The above results show that we obtain improvements if each node either executes a single round of initial estimation or just keeps estimating the stationary distribution while drawing the observations. 
However, it is anticipated that each node suffers a loss in the error exponent if it does not keep estimating the stationary distribution due to the estimation error of the optimal geometric weights. That is, we expect that the dashed lines would eventually divert further from the light blue line (one with the prior knowledge on the stationary distribution) and be outperformed by other solid lines.

\subsection{Quantization}
For each node to obtain the optimal statistic for decision, it could simply disseminate all its observations and likelihood functions into the network. However, the network suffers a high communication cost to support such detailed information flowing among the nodes. The belief-based learning rule, instead, makes each node maintain a real number in binary hypothesis testing. However, a real number is always quantized both being stored at each node or before being transported.

Let us now investigate the effect of quantization through numerical results. Since communication constraints are usually stricter than the computation constraints on each node, we assume that each node is capable of storing a real number while the messages travel among the nodes are quantized.

\begin{figure}[!t]
    \centering
    \includegraphics[width=\linewidth]{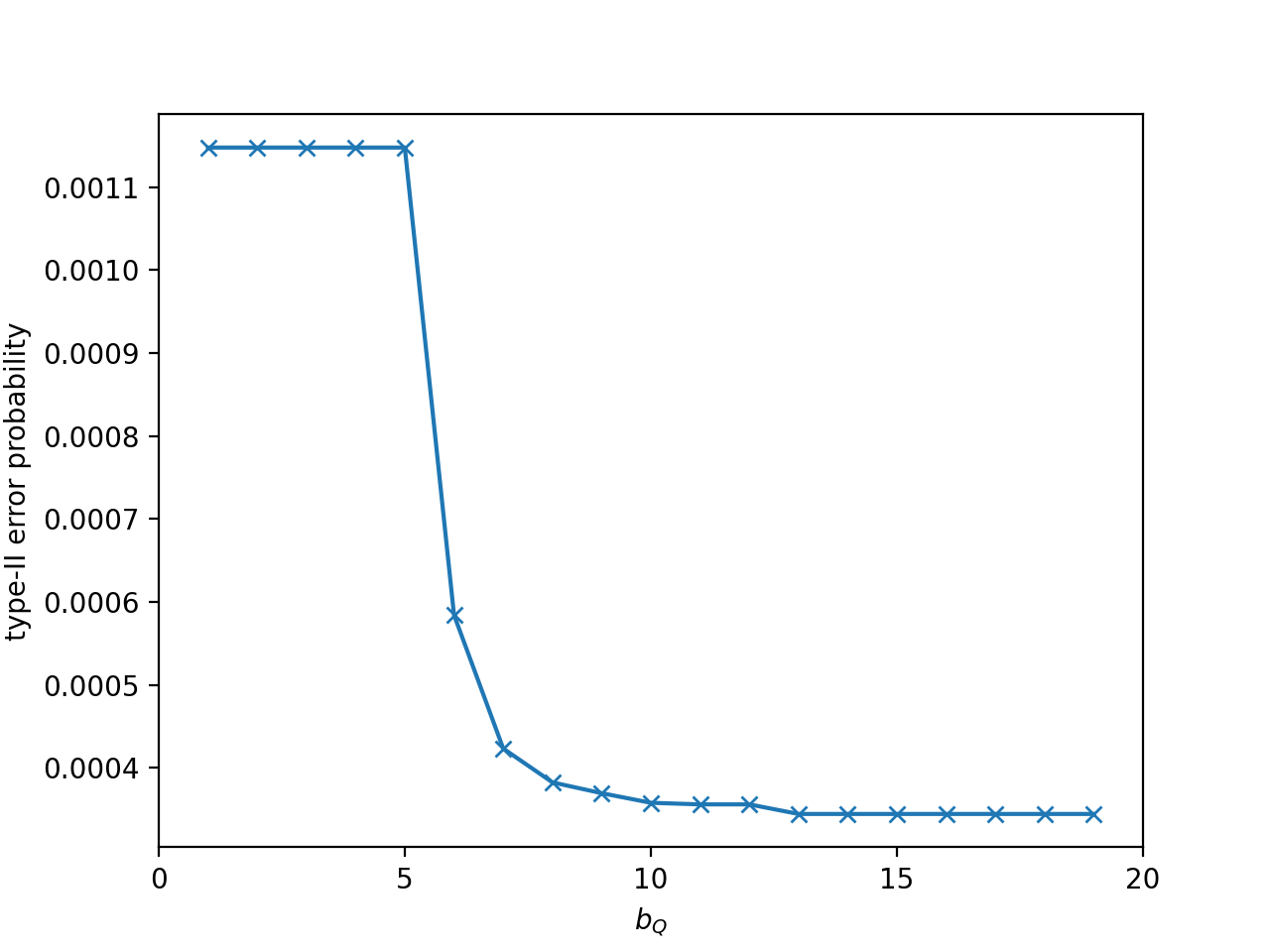}
    \caption{Effect of quantization on the type-II error probability.}
    \label{fig:quantize}
\end{figure}

In the setting for Figure~\ref{fig:quantize}, the network consists of two Bernoulli sources both with parameters $p_0=0.7, p_1=0.8$ under the corresponding two hypotheses $\mathcal{H}_0, \mathcal{H}_1$. The weight matrix is set to be $(W_{11}, W_{12}, W_{21}, W_{22}) = (0.8, 0.2, 0.5, 0.5)$. We consider the Neyman-Pearson problem such that the type-I error probability, $\alpha^{(t)}_i$, is kept under $\epsilon=0.05$ and we use the belief ratio test to obtain the type-II error probability on node~1.

Comparing to the previously proposed learning rule, the main difference in the quantized learning rule is that now the transported messages are quantized. The quantization $Q_b(\cdot)$ transforms the input into its binary representation and keeps the first $b$ bits after the left-most $1$-bit. For example,
\begin{align*}
    Q_3(11_{10}) = Q_3(1011_2) = 1010_2 = 10_{10},
\end{align*}
and
\begin{align*}
    Q_3(0.6875_{10}) = Q_3(0.1011_2) = 0.1010_2 = 0.625_{10}.
\end{align*}
Thus, the final step in the quantized learning rule becomes
\begin{align*}
    \ell^{(t)}_i = W_{ii} \mu^{(t)}_i + \sum_{j \neq i} W_{ij} Q_b(\mu^{(t)}_j).
\end{align*}

Figure~\ref{fig:quantize} shows that the type-II error probability is a lot higher when the communication constraint is stricter and $b \leq 5$. From $b=6$, the error probability drops sharply and soon converges to the optimal type-II error probability under the belief ratio tests.

\section{Discussions and Extensions} \label{sec:discussion}
\subsection{Removing the Assumptions on the Network}\label{subsec:remove_assump}
We assume that the Markov chain governing the information consensus is both irreducible and aperiodic in Assumption~\ref{assum:irreducible_aperiodic}. In this section, we discuss how our learning rule performs without these assumptions.

\begin{figure}[!t]
    \centering
    \includegraphics[width=\linewidth]{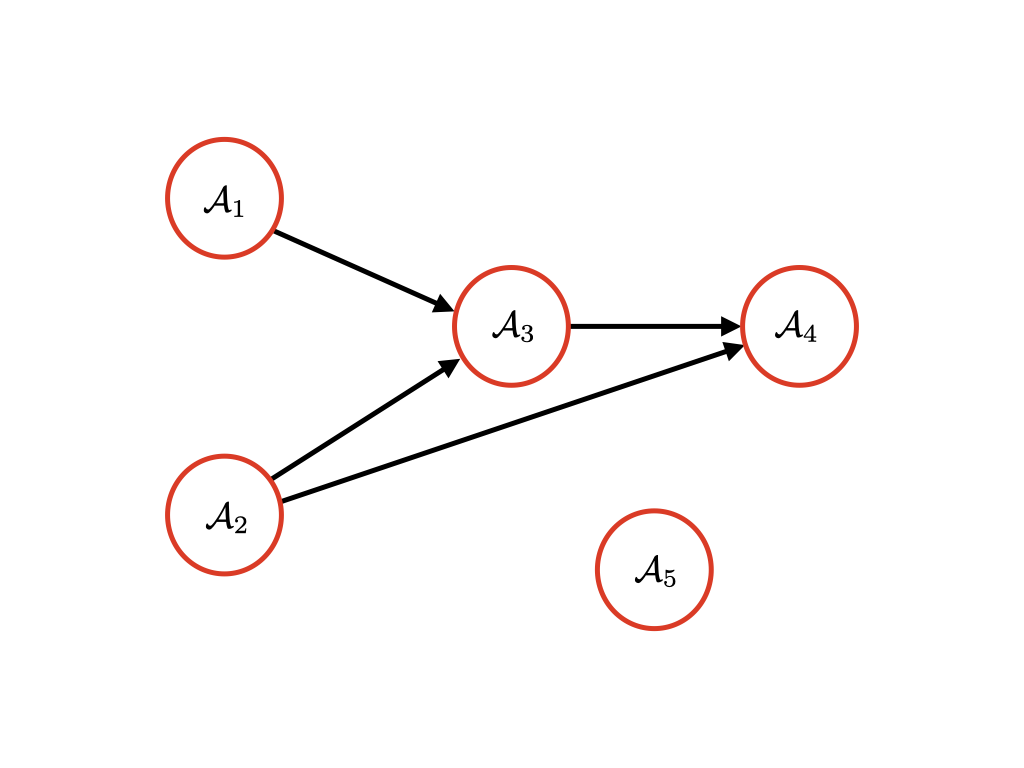}
    \caption{Reducible Markov chain.}
    \label{fig:reducible}
\end{figure}

\paragraph{Irreducible}
For the case that the Markov chain is reducible, we can cluster the states (nodes) into several strongly connected components such that the transition among the components is either unidirectional or none. As illustrated in Figure~\ref{fig:reducible}, the nodes are clustered into $k$ strongly connected components, $\mathcal{A}_1, \dots, \mathcal{A}_4$. The arrow pointing from $\mathcal{A}_1$ to $\mathcal{A}_3$ means that information flows from $\mathcal{A}_3$ to $\mathcal{A}_1$, and formally we have
\begin{align*}
    \exists i \in \mathcal{A}_1 \; \exists j \in \mathcal{A}_3 \; \exists t \in \mathbb{N} \quad [W^t]_{ij} > 0,
\end{align*}
and
\begin{align*}
    \forall i \in \mathcal{A}_1 \; \forall j \in \mathcal{A}_3 \quad W_{ji} = 0.
\end{align*}

It can be shown that we can sort the strongly connected components such that the right ones never point into the left ones such as the network shown in Figure~\ref{fig:reducible}. By rearranging the nodes properly, the weight matrix of the whole network could be written in the following form.
\begin{align*}
    W =
    \begin{bmatrix}
        W_1 & 0 & W_{13} &0 & 0 \\
        0 & W_2 & W_{23} & W_{24} & 0 \\
        0 & 0 & W_3 & W_{34} & 0 \\
        0 & 0 & 0 & W_4 & 0 \\
        0 & 0 & 0 & 0 & W_5
    \end{bmatrix}
\end{align*}

Assume that each component $\mathcal{A}_i$ has size $n_i$. In the weight matrix $W$, each matrix $W_i$ is a square matrix with size $n_i \times n_i$, and the lower-left part of $W$ would be zeros. For $i \neq j$, the matrices $W_{ij}$ represent the edges among the components. It can be shown that the stationary distribution is
\begin{align*}
    \begin{bmatrix}
        \pmb{0} & \pmb{0} & \pmb{0} & \lambda \pi^{(4)} & (1-\lambda) \pi^{(5)}
    \end{bmatrix}
\end{align*}
where $\pi^{(4)}, \pi^{(5)}$ are the unique stationary distributions of $W_4, W_5$, and $\lambda \in [0,1]$. The stationary distribution reveals two problems under our learning rule.

First, we could not set the geometric weights to the inverse of the entries of the stationary distribution since there are zeros in it. Second, the zeros indicate that even if we utilize the original learning rule (without geometric weightings), the information disseminated by the nodes in $\mathcal{A}_1, \mathcal{A}_2$, and $\mathcal{A}_3$ would eventually vanish since the nodes in those components would be overwhelmed by the information coming from $\mathcal{A}_4$. In this case, only the nodes in $\mathcal{A}_4$ and $\mathcal{A}_5$ can pull off the tests.

Thus, our learning rule does not work under an irreducible network. However, if the inter-component edges are controlled by some routers, the problem might be overcome.

\paragraph{Aperiodic} We consider the case where the Markov chain consists of a single strongly connected component. The period of a state (node) is
\begin{align*}
    T_i = \mathrm{gcd} \left\{ t \in \mathbb{N} : [W^t]_{ii} > 0 \right\},
\end{align*}
and node $i$ is periodic with period $T_i$ if $T_i>1$. If a node is periodic with period $T$, then the other nodes in the same strongly connected component have the same period $T$. For convenience, we say that the component has period $T$.

\begin{figure}[!t]
     \centering
     \includegraphics[width=\linewidth]{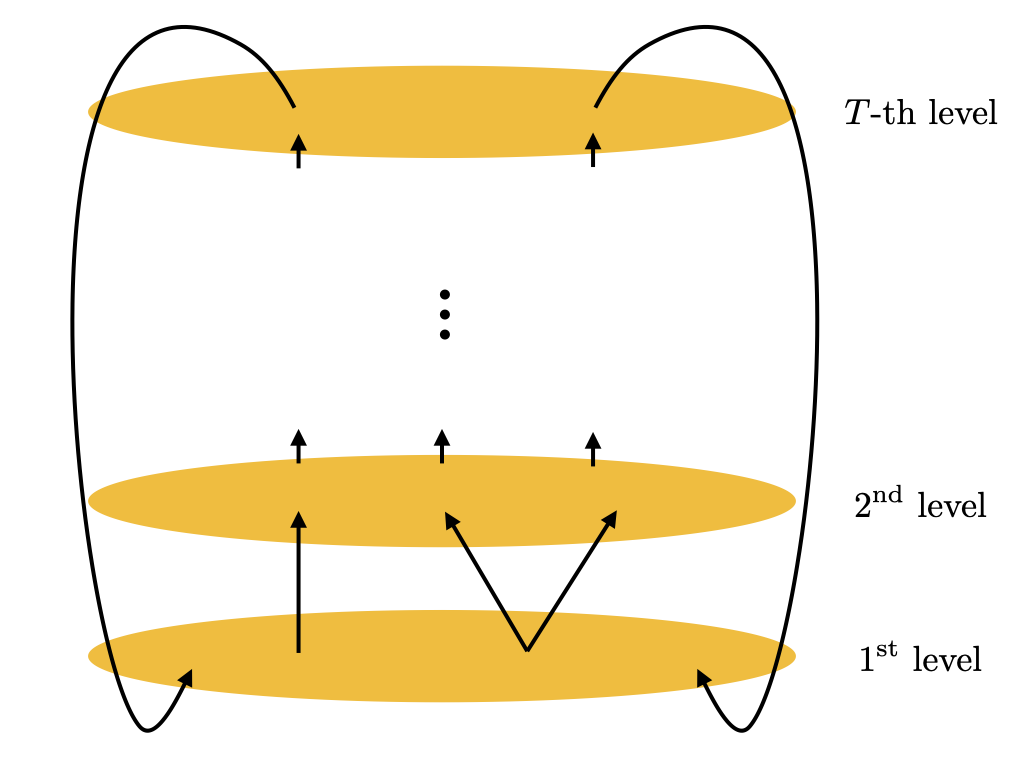}
     \caption{Sorting nodes into levels for a periodic network.}
     \label{fig:periodic_level}
\end{figure}

For a strongly connected component with period $T$, we can sort the nodes into $T$ levels in Figure~\ref{fig:periodic_level} such that each node in the $v$-th level points toward the nodes in the $(v+1)$-th level, and the nodes in the last level point back to the ones in the first level. For any observation drawn by a node at time $t$, the piece of information (the log-likelihood ratio) disseminated by the node flows through the levels and return to the node at time $t+T$. Thus, asymptotically, each node has only $\frac{1}{T}$ of the total number of pieces of information, and the error exponent on each node is
\begin{align*}
    \lim_{t \rightarrow \infty} - \frac{1}{t} \log \beta^{(t)*}_i(r^*; \epsilon) = \frac{1}{T} \sum_{j=1}^n D_{\mathrm{KL}} \left( P_{j,0} \Vert P_{j,1} \right).
\end{align*}

\subsection{Multiple Hypothesis Testing}
\label{sec:multiple_hypothesis_testing}
In our main results, we consider the binary hypothesis testing problem. For the $M$-ary hypothesis testing, as we mentioned in Remark~\ref{rem:geometric_weight}, we can modify the second step in the original learning rule in~\cite{lalitha2018social} into:
\begin{align*}
    b_{i}^{(t)}(\theta) = \frac{ \left( P_{i,\theta}(X_i^{(t)}) \right)^{r_i} q_{i}^{(t-1)}(\theta) }{ \sum_{a=1}^M \left( P_{i,a}(X_i^{(t)}) \right)^{r_i} q_{i}^{(t-1)}(a) }.
\end{align*}
By choosing the geometric weights as the inverse of the stationary distribution, the belief ratio again mimics the likelihood ratio and could be utilized for the ratio test. In this case, for example, the error exponent of the Bayes risk would be $\mathsf{P}^{(t)*}_{\mathsf{e},i}(r^*; \xi) = \min_{j \neq k} \mathrm{CI}(P_j, P_k)$.

For multiple hypothesis testing with a rejection option, to keep the rejection probability under a certain constant level, we consider the following decision rule for each node $i \in [n]$.
\begin{itemize}
    \item If $\log \frac{b^{(t)}_i(\theta_k)}{b^{(t)}_i(\theta_l)} \geq \gamma^*_{i}(\theta_k, \theta_l)$ for all $l \neq k$, then $\hat{\theta}_i = \theta_k$.
    \item If the test failed for all $k$, then $\hat{\theta}_i = \mathsf{R}$.
\end{itemize}
Each threshold $\gamma^*_{k,l}$ is chosen such that
\begin{align*}
     &P_k \left\{ \log \frac{b^{(t)}_i(\theta_k)}{b^{(t)}_i(\theta_l)} < \gamma^*_{i}(\theta_k, \theta_l) \right\}
    < \epsilon\quad \text{and}\\
     &P_k \left\{ \log \frac{b^{(t)}_i(\theta_k)}{b^{(t)}_i(\theta_l)} \leq \gamma^*_{i}(\theta_k, \theta_l) \right\} \ge \epsilon.
\end{align*}
Then the probability of rejection for each node $i$ under $\theta_k$ is
\begin{align*}
    \mathsf{P}_{\mathsf{R},i}
    =&\; 1 - P_k \left\{ \exists m \; \forall l \neq m :\; \log \frac{b^{(t)}_i(\theta_m)}{b^{(t)}_i(\theta_l)} \geq \gamma^*_{i}(\theta_m, \theta_l) \right\} \\
    \leq&\; 1 - P_k \left\{ \forall l \neq k :\; \log \frac{b^{(t)}_i(\theta_k)}{b^{(t)}_i(\theta_l)} \geq \gamma^*_{i}(\theta_k, \theta_l) \right\} \\
    =&\; P_k \left\{ \exists l \neq k :\; \log \frac{b^{(t)}_i(\theta_k)}{b^{(t)}_i(\theta_l)} < \gamma^*_{i}(\theta_k, \theta_l) \right\} \\
    \leq&\; \sum_{l \neq k} P_k \left\{ \log \frac{b^{(t)}_i(\theta_k)}{b^{(t)}_i(\theta_l)} < \gamma^*_{i}(\theta_k, \theta_l) \right\} \\
    \leq&\; (M-1) \epsilon,
\end{align*}
which is controlled by the constant $\epsilon$. If we choose the thresholds with $ \epsilon = \frac{\epsilon'}{M-1}$, we get the probability of rejection upper bounded by $\epsilon'$.

Meanwhile, on node $i$, each probability of error is
\begin{align*}
    P_k \left\{ \hat{\theta}_i = \theta_l \right\}
    &\leq P_k \left\{ \log \frac{b^{(t)}_i(\theta_k)}{b^{(t)}_i(\theta_l)} < \gamma^*_{i}(\theta_l, \theta_k   ) \right\}\\
    &\doteq e^{- tD_{\mathrm{KL}}(P_l \Vert P_k)}
\end{align*}
by our theorems for the binary case in previous sections. Thus, we can see that our results could be extended to these scenarios.

\section{Conclusion} \label{sec:conclusion}
In this work, we study the price of decentralization in distributed hypothesis testing. The original learning rule introduced in Section~\ref{sec:social_learning_rule} obtains a tilted statistic compared to the centralized one and thus leads to a sub-optimal error exponent. The sub-optimal result comes from the network imbalance, and we compensate for the imbalance by properly choosing the additional geometric weights introduced in our modified learning rule to achieve the optimal error exponent. Furthermore, we look for the higher-order asymptotics of the type-II error probability and the Bayes risk. We reveal an upper bound on the price of decentralization as a constant term in the exponent of the error probability, where the extra term depends on the connectivity of the underlying network and the network imbalance.

We propose an estimation rule for obtaining the geometric weight on each node and form a combined learning rule. Simulation results support the relationship between the probability of error and the network imbalance and show how much improvement our learning rule obtains in terms of the probability of error. The simulation of the quantized learning rule gives us a glimpse of the effect of quantization. Other discussions, extensions, and future work are also shown.

Some directions are left as future work. First, our first-order results are optimal, while our analysis on the higher-order terms turns out to be upper bounds on the error probability. Whether a constant penalty is inevitable (except for a complete graph with uniform weights) remains unclear, and it is anticipated that a non-trivial lower bound on the error probability is needed to resolve the question. Secondly, despite the promising simulation results shown in Section~\ref{sec:simulations}, a rigorous analysis on the probability of error while estimating the unknown geometric weight is lacking. Last but not least, taking the communication cost into account is an important next step towards a refined understanding of the price of decentralization in decentralized learning in practice.

\appendix
\section{Proofs}\label{sec:proofs}
\subsection{Proof of Theorem~\ref{thm:decentralized_optmization_form}}
\label{pf:decentralized_optmization_form}

Under the log-belief ratio test, we can obtain upper bounds on both the type-I and type-II error probability. For each node $i \in [n]$, we have
\begin{align*}
\alpha_i^{(t)}(r; \eta_i^{(t)}, \gamma_i^{(t)})
&= P_0 \left\{  \phi_i^{(t)}(\ell_i^{(t)}) = 1 \right\}\\
&= P_0 \left\{ \ell_i^{(t)} > \gamma_i^{(t)} \right\} + \eta_i^{(t)} P_0 \left\{ \ell_i^{(t)} = \gamma_i^{(t)} \right\}\\
&\leq P_0 \left\{ \ell_i^{(t)} \geq \gamma_i^{(t)} \right\},
\end{align*}
and
\begin{align}
    \beta_i^{(t)}(r; \eta_i^{(t)}, \gamma_i^{(t)})
    &= P_1 \left\{  \phi_i^{(t)}(\ell_i^{(t)}) = 1 \right\} \notag\\
    &\leq P_1 \left\{ \ell_i^{(t)} \leq \gamma_i^{(t)} \right\}, \label{eq:beta_upp}
\end{align}
where to simplify the notations, we use 
\begin{align*}
    P_\theta \left\{ E \right\} \triangleq \mathrm{Pr} \left\{ E \;|\; \mathcal{H}_\theta \right\}    
\end{align*}
to denote the probability of event $E$ occurring given $\mathcal{H}_\theta$ is the true hypothesis. 

Given the geometric weights, $r_i$'s, we can recursively decompose the log-belief by the learning rule ratio as
\begin{align*}
&\ell_i^{(t)}\\
&= \sum_{j=1}^n W_{ij} \mu_j^{(t)}
= \sum_{j=1}^n W_{ij} r_j \log \frac{ P_{j,1}(X_j^{(t)})}{ P_{j,0}(X_j^{(t)}) } + \sum_{j=1}^n W_{ij} \ell_j^{(t-1)} \\
&= \sum_{j=1}^n \sum_{\tau=1}^t [W^\tau]_{ij} r_j \log \frac{ P_{j,1}(X_j^{(t-\tau+1)}) }{ P_{j,0}(X_j^{(t-\tau+1)}) } + \sum_{j=1}^n \left[ W^t \right]_{ij} \log \ell_j^{(0)} \\
&= \sum_{j=1}^n \sum_{\tau=1}^t [W^\tau]_{ij} r_j \log \frac{ P_{j,1}(X_j^{(t-\tau+1)}) }{ P_{j,0}(X_j^{(t-\tau+1)}) },
\end{align*}
where we use $\left[ W^t \right]_{ij}$ to denote the $(i,j)$-th entry of $W^t$. 

Since the observations are mutually independent, we can now further write~\eqref{eq:beta_upp} as
\begin{align}
&\beta_i^{(t)}(r; \eta^{(t)}_i, \gamma^{(t)}_i) \notag\\
&\leq P_1 \left\{ \sum_{j=1}^n \sum_{\tau=1}^t [W^\tau]_{ij} r_j \log \frac{ P_{j,1}(X_j^{(t-\tau+1)}) }{ P_{j,0}(X_j^{(t-\tau+1)}) } \leq \gamma_i^{(t)} \right\} \nonumber \\
&= P_1 \left\{ \sum_{j=1}^n \sum_{\tau=1}^t [W^\tau]_{ij} r_j \log \frac{ P_{j,1}(X_j^{(\tau)}) }{ P_{j,0}(X_j^{(\tau)}) } \leq \gamma_i^{(t)} \right\}. \label{eq:beta_upp_2}
\end{align}

To bound the probability in~\eqref{eq:beta_upp_2}, we introduce a large deviation analysis. For simplicity, let the random vector $Y_i^{(t)}$ be
\begin{align*}
Y_i^{(t)} &\triangleq \left[ Y_{i1}^{(t)} \; \dots \; Y_{in}^{(t)} \right]^{\mathrm{T}}, \\
Y_{ij}^{(t)} &\triangleq \left[ W^t \right]_{ij} r_j \log \frac{ P_{j,1}(X_j^{(t)}) }{ P_{j,0}(X_j^{(t)}) },
\end{align*}
and since $\lim_{t \rightarrow \infty} \left[ W^t \right]_{ij} = \pi_j$, let
\[
Y \triangleq \left[ Y_1 \; \dots \;Y_n \right]^{\mathrm{T}}, \ 
Y_j \triangleq \lim_{t \rightarrow \infty} Y_{ij}^{(t)} \sim \pi_j r_j \log \frac{ P_{j,1}(X_j) }{ P_{j,0}(X_j) }.
\]
Furthermore, let $Z_{ij}^{(t)}$ be the empirical mean of $Y_{ij}^{(t)}$ such that
\[
Z_i^{(t)} \triangleq \left[ Z_{i1}^{(t)} \; \dots \; Z_{in}^{(t)} \right]^{\mathrm{T}},\ 
Z_{ij}^{(t)} \triangleq \frac{1}{t} \sum_{\tau=1}^t Y_{ij}^{(\tau)}.
\]
Substitute the definitions into~\eqref{eq:beta_upp_2}, we have
\begin{align*}
\beta_i^{(t)}(r; \eta^{(t)}_i, \gamma^{(t)}_i)
&\leq P_1 \left\{ \sum_{j=1}^n \sum_{\tau=1}^t Y_{ij}^{(t)} \leq \gamma_i^{(t)} \right\}\\
&= P_1 \left\{ \sum_{j=1}^n Z_{ij}^{(t)} \leq \frac{1}{t} \gamma_i^{(t)} \right\}.
\end{align*}
Define the logarithmic moment generating function of $Z_i^{(t)}$, $\Lambda_t: \mathbb{R}^n \rightarrow \mathbb{R}$, as
\begin{align*}
    \Lambda_t(\lambda) \triangleq \log \mathsf{E}_{P_1} \left[ e^{ \langle \lambda, Z_i^{(t)} \rangle } \right],
\end{align*}
and define
\begin{align}
    \Lambda(\lambda)
    &\triangleq \lim_{t \rightarrow \infty} \frac{1}{t} \Lambda_t(t \lambda)
    = \lim_{t \rightarrow \infty} \frac{1}{t} \log \mathsf{E}_{P_1} \left[ e^{ \langle t \lambda, Z_i^{(t)} \rangle } \right] \notag\\
    &= \lim_{t \rightarrow \infty} \frac{1}{t} \log \mathsf{E}_{P_1} \left[ e^{ \langle \lambda, \sum_{\tau=1}^t Y_i^{(\tau)} \rangle } \right] \nonumber \\
    &= \lim_{t \rightarrow \infty} \frac{1}{t} \log \left( \prod_{\tau=1}^t \mathsf{E}_{P_1} \left[ e^{ \langle \lambda, Y_i^{(\tau)} \rangle } \right] \right) \notag\\
    &= \lim_{t \rightarrow \infty} \frac{1}{t} \sum_{\tau=1}^t \log \mathsf{E}_{P_1} \left[ e^{ \langle \lambda, Y_i^{(\tau)} \rangle } \right], \label{eq:lmgf_1}
\end{align}
where the subscript $P_1$ in the expectation denotes that the expectation is taken over the distribution under $\mathcal{H}_1$. 

Since $\lim_{t \rightarrow \infty} Y_i^{(t)} \sim Y$, we have
\begin{align}
    \lim_{t \rightarrow \infty} \log \mathsf{E}_{P_{1}} \left[ e^{ \langle \lambda, Y_i^{(t)} \rangle } \right] =\log \mathsf{E}_{P_{1}} \left[ e^{ \langle \lambda, Y \rangle } \right], \nonumber
\end{align}
and by recognizing~\eqref{eq:lmgf_1} as the Cesàro summation of the series $\sum_{\tau=1}^t \log \mathsf{E}_{P_{1}} \left[ e^{ \langle \lambda, Y_i^{(\tau)} \rangle } \right]$, we have
\begin{align}
    \Lambda(\lambda) = \log \mathsf{E}_{P_{1}} \left[ e^{ \langle \lambda, Y \rangle } \right]. \nonumber
\end{align}
The Fenchel-Legendre transform of $\Lambda(\lambda)$ is
\begin{align}
    \Lambda^*(x) \triangleq \sup_{\lambda \in \mathbb{R}^n} \left\{ \langle \lambda, x \rangle - \Lambda(\lambda) \right\}. \nonumber 
\end{align}

To this end, let us recall the Gärtner-Ellis Theorem from the large deviation theory as follows.
\begin{lemma}[Gärtner-Ellis Theorem~\cite{dembo_zeitouni_2010}] \label{lem:GE}
Consider a sequence of random vector $Z^{(t)} \in \mathbb{R}^n$ with law $\mu_t$ and logarithmic moment generating function $\Lambda_t(\lambda)$. If the limit
\begin{align*}
    \Lambda(\lambda) = \lim_{t \rightarrow \infty} \frac{1}{t} \Lambda_t(t \lambda)
\end{align*}
exists as an extended real number and the origin belongs to the interior of $D_{\Lambda} \triangleq \left\{ \lambda \in \mathbb{R}^n: \Lambda(\lambda) < \infty \right\}$, for any closed set $\mathcal{F}$
\begin{align*}
    \liminf_{t \rightarrow \infty} - \frac{1}{t} \log \mu_t(\mathcal{F}) \geq \inf_{x \in \mathcal{F}} \Lambda^*(x).
\end{align*}
\end{lemma}

By the weak law of large numbers, if we let
\begin{align}
    \gamma_i^{(t)} = t \left( \sum_{j=1}^n \pi_j r_j \mathsf{E}_{P_0} \left[ \log \frac{P_{j,1}(X_j)}{P_{j,0}(X_j)} \right] + \delta \right) \nonumber
\end{align}
for some $\delta>0$, we have $\alpha_i^{(t)}(r; \eta^{(t)}_i, \gamma^{(t)}_i) < \epsilon$ for all $i$ for $n$ sufficiently large. 

Let $\gamma_i = \frac{1}{t} \gamma_i^{(t)}$ and the closed set $\mathcal{F} \subset \mathbb{R}^n$ be
\begin{align*}
    \mathcal{F} = \left\{ x \in \mathbb{R}^n: \sum_{i=1}^n x_i \leq \gamma_i. \right\}.
\end{align*}
By Lemma~\ref{lem:GE}, we have
\begin{align}
    &\liminf_{t \rightarrow \infty} - \frac{1}{t} \log \beta_i^{(t)}(r; \eta^{(t)}_i, \gamma^{(t)}_i) \notag\\
    &\geq \liminf_{t \rightarrow \infty} - \frac{1}{t} \log P_1 \left\{ \sum_{j=1}^n Z_{ij}^{(t)} \leq \gamma_i \right\} \notag\\
    &\geq \inf_{x \in \mathcal{F}} \Lambda^*(x) \nonumber \\
    &= \inf_{x \in \mathcal{F}} \sup_{\lambda \in \mathbb{R}^n} \left\{ \langle \lambda, x \rangle - \Lambda(\lambda) \right\} \nonumber \\
    &= \sup_{\lambda \in \mathbb{R}^n} \inf_{x \in \mathcal{F}} \left\{ \langle \lambda, x \rangle - \log \mathsf{E}_{P_1} \left[ e^{\langle \lambda, Y \rangle} \right] \right\} \label{eq:sup_inf} \\
    &= \sup_{\lambda \in \mathbb{R}_{\leq 0}^n} \inf_{x \in \mathcal{F}} \left\{ \langle \lambda, x \rangle - \log \mathsf{E}_{P_1} \left[ e^{\langle \lambda, Y \rangle} \right] \right\} \label{eq:lam_leq0} \\
    &= \sup_{\lambda \in \mathbb{R}_{\leq 0}} \inf_{x \in \mathcal{F}} \left\{ \langle \lambda \pmb{1}, x \rangle - \log \mathsf{E}_{P_1} \left[ e^{\langle \lambda \pmb{1}, Y \rangle} \right] \right\} \label{eq:lam_iden} \\
    &= \sup_{\lambda \in \mathbb{R}_{\leq 0}} \left\{ \langle \lambda\pmb{1}, \frac{\gamma_i}{n} \pmb{1} \rangle - \log \mathsf{E}_{P_1} \left[ e^{\lambda \langle \pmb{1}, Y \rangle} \right] \right\} \nonumber \\
    &= \sup_{\lambda \in \mathbb{R}_{\leq 0}} \left\{ \lambda \gamma_i - \log \mathsf{E}_{P_1} \left[ e^{\lambda \sum_{j=1}^n Y_j} \right] \right\}. \nonumber
\end{align}
Here $\mathbb{R}_{\leq 0}$ denotes the set of non-positive real numbers. 
Since the logarithmic moment generating function $\Lambda(\lambda)$ is convex in $\lambda$, by the minimax theorem, we exchange the order of the infimum and supremum in~\eqref{eq:sup_inf}. In~\eqref{eq:lam_leq0}, observe that if $\lambda$ has any positive entry, the term $\langle \lambda, x \rangle$ easily goes to negative infinity, and thus the optimal $\lambda$ must fall in $\mathbb{R}_{\leq0}^n$. Furthermore, if $\lambda$ is not orthogonal to the boundary of $\mathcal{F}$, that is, $x_1 + \dots + x_n = \gamma_i$, we can always find an $x \in \mathcal{F}$ such that $\langle \lambda, x \rangle$ goes to negative infinity. Let $\pmb{1} \in \mathbb{R}^n$ denote the vector with all entries being 1. Thus in~\eqref{eq:lam_iden} we simplify the optimization problem Instead of optimizing over $\lambda \in \mathbb{R}_{\leq0}^n$. Now we can see that $x^*$ is optimal if and only if $x^*_1 + \dots + x^*_n = \gamma_i$, thus we choose $x = \frac{\gamma_i}{n} \pmb{1}$ and plug it in into~\eqref{eq:lam_iden}. Plug the $\gamma_i$ we chose earlier in, and we have
\begin{align}
&\sup_{\lambda \leq 0} \left\{ \lambda \gamma_i - \log \mathsf{E}_{P_1} \left[ e^{\lambda \sum_{j=1}^n Y_j} \right] \right\} \notag \\
&= \sup_{\lambda \leq 0} \left\{
\begin{array}{l}
\lambda \left( \sum_{j=1}^n \pi_j r_j \mathsf{E}_{P_0} \left[ \log \frac{P_{j,1}(X_j)}{P_{j,0}(X_j)} \right] + \delta \right) \\
 - \sum_{j=1}^n \log \mathsf{E}_{P_1} \left[ e^{\lambda Y_j} \right]
\end{array} 
\right\} \nonumber \\
&= \sup_{\lambda \leq 0} \left\{
\begin{array}{l}
\lambda \delta + \sum_{j=1}^n \lambda \pi_j r_j \mathsf{E}_{P_0} \left[ \log \frac{P_{j,1}(X_j)}{P_{j,0}(X_j)} \right] \\
- \log \mathsf{E}_{P_1} \left[ \exp \left( \lambda \pi_j r_j \log \frac{P_{j,1}(X_j)}{P_{j,0}(X_j)} \right) \right]
\end{array} \right\} \nonumber \\
&= \sup_{\lambda \geq 0} \left\{
\begin{array}{l}
- \lambda \delta + \sum_{j=1}^n \lambda \pi_j r_j \mathsf{E}_{P_0} \left[ \log \frac{P_{j,0}(X_j)}{P_{j,1}(X_j)} \right] \\
- \log \mathsf{E}_{P_1} \left[ \left( \frac{P_{j,0}(X_j)}{P_{j,1}(X_j)} \right)^{\lambda \pi_j r_j} \right]
\end{array} \right\}. \label{eq:lam_opt}
\end{align}
Since we can make $\delta$ arbitrarily close to zero, we can omit the first term in~\eqref{eq:lam_opt} and we have
\begin{align}
&\liminf_{t \rightarrow \infty} \frac{1}{t} \beta^{(t)*}_i(r, \epsilon) \notag\\
&\geq \sup_{\lambda \geq 0} \left\{ 
\begin{array}{l}\sum_{j=1}^n \lambda \pi_j r_j \mathsf{E}_{P_0} \left[ \log \frac{P_{j,0}(X_j)}{P_{j,1}(X_j)} \right]\\ - \log \mathsf{E}_{P_1} \left[ \exp \left( \lambda \pi_j r_j \log \frac{P_{j,0}(X_j)}{P_{j,1}(X_j)} \right) \right] 
\end{array}\right\}. \label{eq:lam_opt2}
\end{align}

The upper bound on the error exponent is obtained by the other part of the Gärtner-Ellis Theorem~\cite{dembo_zeitouni_2010} with a similar technique and hence omitted here.

\subsection{Proof of Theorem~\ref{thm:neyman_pearson_exponent}}
\label{pf:neyman_pearson_exponent}
Choose the geometric weights as
\[
r_i = \frac{c}{\pi_i} \quad \forall i \in [n]
\]
with any constant $c > 0$. 

To find the optimal $\lambda$ in~\eqref{eq:lam_opt2}, we set the derivative over $\lambda$ to zero, that is,
\begin{align*}
0 &= \sum_{j=1}^n c \, \mathsf{E}_{P_0} \left[ \log \frac{P_{j,0}(X_j)}{P_{j,1}(X_j)} \right] \\
&\quad - \sum_{j=1}^n\frac{ \mathsf{E}_{P_1} \left[ c \left( \log \frac{P_{j,0}(X_j)}{P_{j,1}(X_j)} \right) \left( \frac{P_{j,0}(X_j)}{P_{j,1}(X_j)} \right)^{\lambda c} \right] }{ \mathsf{E}_{P_1} \left[ \left( \frac{P_{j,0}(X_j)}{P_{j,1}(X_j)} \right)^{\lambda c} \right] }.
\end{align*}
Plug in $\lambda = c^{-1}$, the above is satisfied because
\begin{align*}
&\frac{ c \,\mathsf{E}_{P_1} \left[ \left( \log \frac{P_{j,0}(X_j)}{P_{j,1}(X_j)} \right) \left( \frac{P_{j,0}(X_j)}{P_{j,1}(X_j)} \right)^{\lambda c} \right] }{ \mathsf{E}_{P_1} \left[ \left( \frac{P_{j,0}(X_j)}{P_{j,1}(X_j)} \right)^{\lambda c} \right] } \\
&=  \frac{ \mathsf{E}_{P_1} \left[ c \left( \log \frac{P_{j,0}(X_j)}{P_{j,1}(X_j)} \right) \left( \frac{P_{j,0}(X_j)}{P_{j,1}(X_j)} \right)  \right] }{ \mathsf{E}_{P_1} \left[ \frac{P_{j,0}(X_j)}{P_{j,1}(X_j)} \right] } \\
&= \mathsf{E}_{P_0} \left[ c \log \frac{P_{j,0}(X_j)}{P_{j,1}(X_j)} \right] .
\end{align*}
Since the term in the supremum of~\eqref{eq:lam_opt2} is concave in $\lambda$, thus $\lambda = c^{-1}$ must be the optimal solution of $\lambda$ if we choose $r_i = c / \pi_i$ for all $i \in [n]$. Denote the choice of such an $r$ as $\pi^{-1}$, we have
\begin{align*}
&\liminf_{t\rightarrow \infty} - \frac{1}{t} \log \beta_i^{(t)*}(\pi^{-1}, \epsilon)\\
&\geq \sum_{j=1}^n \mathsf{E}_{P_0} \left[ \log \frac{P_{j,0}(X_j)}{P_{j,1}(X_j)} \right] - \log \mathsf{E}_{P_1} \left[ \frac{P_{j,0}(X_j)}{P_{j,1}(X_j)} \right]\\
&= \sum_{j=1}^n D_{\mathrm{KL}} \left( P_{j,0} \Vert P_{j,1} \right).
\end{align*}
Since the convergence rate in the decentralized regime cannot outperform the rate in the centralized regime, we must have
\begin{align*}
\limsup_{t\rightarrow \infty} - \frac{1}{t} \log \beta_i^{(t)*}(\pi^{-1}, \epsilon) \leq \sum_{j=1}^n D_{\mathrm{KL}} \left( P_{j,0} \Vert P_{j,1} \right)
\end{align*}
and Theorem~\ref{thm:neyman_pearson_exponent} is proved.

\subsection{Proof of Theorem~\ref{thm:bayes_exponent}}
\label{pf:bayes_exponent}
We show that by setting the threshold of log-belief ratio test to zero and choosing $r=\pi^{-1}$, both type-I and type-II error have the same convergence rate which is the Chernoff information over the nodes' product distribution. Recall that
\begin{gather*}
\alpha_i^{(t)}(r; \eta^{(t)}_i, \gamma^{(t)}_i) \leq P_0 \Bigg\{ \sum_{j=1}^n Z_j^{(t)} \geq \gamma_i^{(t)} \Bigg\}, \\
\beta_i^{(t)}(r; \eta^{(t)}_i, \gamma^{(t)}_i) \leq P_1 \Bigg\{ \sum_{j=1}^n Z_j^{(t)} \leq \gamma_i^{(t)} \Bigg\}.\end{gather*}
By Lemma~\ref{lem:GE}, we have
\begin{align*}
& \liminf_{t \rightarrow \infty} - \frac{1}{t} \log \alpha_i^{(t)}(r; \eta^{(t)}_i, \gamma^{(t)}_i)\\
&\geq \liminf_{t \rightarrow \infty} - \frac{1}{t} \log P_0 \Bigg\{ \sum_{j=1}^n Z_j^{(t)} \geq \gamma_i^{(t)} \Bigg\}\\
&\geq \inf_{x \in \mathcal{F}} \Lambda^*(x) \\
&= \inf_{x \in \mathcal{F}} \sup_{\lambda \in \mathbb{R}^n} \big\{ \langle \lambda, x \rangle - \Lambda(\lambda) \big\}\\
&= \sup_{\lambda \geq 0} \Bigg\{ \lambda \gamma_i^{(t)} - \sum_{j=1}^n \log \mathsf{E}_{P_0} \Bigg[ \exp \Bigg( \lambda \pi_j r_j \log \frac{P_{j,1}(X_j)}{P_{j,0}(X_j)} \Bigg) \Bigg] \Bigg\}.
\end{align*}
Choose $r = \pi^{-1}$ and let the threshold in the log-belief ratio test $\gamma_i^{(t)}$ be zero. Then, we have
\begin{align*}
&\liminf_{t \rightarrow \infty} - \frac{1}{t} \log \alpha_i^{(t)}(\pi^{-1}; \eta^{(t)}_i, 0)\\
&\geq \sup_{\lambda \geq 0} \Bigg\{ - \sum_{j=1}^n \log \mathsf{E}_{P_0} \Bigg[ \exp \Bigg( \lambda \log \frac{P_{j,1}(X_j)}{P_{j,0}(X_j)} \Bigg) \Bigg] \Bigg\} \\
&= \sup_{\lambda \geq 0} \Bigg\{ - \sum_{j=1}^n \log \Bigg( \int_{x \in \mathcal{X}_j} P_{j,0}(x)^{1-\lambda} P_{j,1}(x)^{\lambda} dx \Bigg) \Bigg\} \\
&= \sup_{\lambda \geq 0} \Bigg\{ - \log \Bigg( \int_{x \in \mathcal{X}} P_0(x)^{1-\lambda} P_1(x)^{\lambda} dx \Bigg) \Bigg\} \\
&= \mathrm{CI}(P_0, P_1),
\end{align*}
where $\mathcal{X} = (\mathcal{X}_1, \dots, \mathcal{X}_n)$ and $P_{\theta}$ is the product distribution of $P_{1,\theta}, \dots, P_{n,\theta}$ for all $\theta \in \{0,1\}$.
Through a similar derivation, we also have
\begin{align}
    \liminf_{t  \rightarrow \infty} - \frac{1}{t} \log \beta_i^{(t)}(\pi^{-1}; \eta^{(t)}_i, 0) \geq \mathrm{CI}(P_0, P_1). \nonumber
\end{align}
Since the convergence rate must not outperform the one in the centralized regime, Theorem~\ref{thm:bayes_exponent} is proved.

\subsection{Proof of Theorem~\ref{thm:neyman_pearson_higher_order}}
\label{pf:neyman_pearson_higher_order}

Let $X = \left( X_1^{(1)}, X_2^{(1)}, \dots, X_n^{(t)} \right)$ denote the sequence of all observations in the first $t$ rounds, $\mathcal{X}^t = (\mathcal{X}_1, \dots, \mathcal{X}_n)^{\otimes t}$ denote the product sample space, and
\begin{align*}
h^{(t)}(X) &= \sum_{\tau=1}^t \sum_{j=1}^n \log \frac{ P_{j,1}(X_j^{(\tau)}) }{ P_{j,0}(X_j^{(\tau)}) }, \\
\tilde{h}^{(t)}_i ( X ) &= \sum_{\tau=1}^t \sum_{j=1}^n \frac{\left[ W^\tau \right]_{ij}}{\pi_j}\log \frac{ P_{j,1}(X_j^{(\tau)}) }{ P_{j,0}(X_j^{(\tau)}) }.
\end{align*}
For $X_j \sim P_{j,0}$, we have
\begin{align*}
H_n &= \frac{1}{t} \mathsf{E}_{P_0}\left[ h^{(t)}(X) \right] = - \sum_{j=1}^n \mathrm{D}\left( P_{j,0} \Vert P_{j,1} \right), \\
\tilde{H}_i^{(t)} &= \frac{1}{t} \mathsf{E}_{P_0}\left[ \tilde{h}^{(t)}_i(X) \right]
= - \frac{1}{t} \sum_{\tau=1}^t \sum_{j=1}^n \frac{\left[ W^\tau \right]_{ij}}{\pi_j} \mathrm{D}\left( P_{j,0} \Vert P_{j,1} \right)
\end{align*}
for all $i \in [n]$.  
Furthermore, let $S_n^2$ and $\alpha_n$ denote the second and third central moment of
\begin{align*}
\sum_{j=1}^n \log \frac{ P_{j,1}(X_j) }{ P_{j,0}(X_j) },
\end{align*}
where $X_j \sim P_{j,0}$ for all $j \in [n]$. 
Let $\gamma_i^{(t)}$ be the threshold such that the type-I error at time $t$ is less than or equal to $\epsilon$, then
\begin{align}
&\beta_i^{(t)}(r^*; \eta^{(t)}_i, \gamma^{(t)}_i) \notag\\
&\leq P_1 \left\{ \ell^{(t)}_i \leq \gamma_i^{(t)} \right\} \nonumber \\
&= \sum_{x \in \mathcal{X}^t} P_1^{\otimes t}(x) \mathbbm{1} \left\{ \sum_{\tau=1}^t \sum_{j=1}^n \frac{\left[ W^\tau \right]_{ij}}{\pi_j} \log \frac{P_{j,1}(x_j^{(\tau)})}{P_{j,0}(x_j^{(\tau)})} \leq \gamma_i^{(t)} \right\} \nonumber \\
&= 
\sum_{x \in \mathcal{X}^t}\left\{
\begin{array}{l} P_0^{\otimes t}(x) \exp \left\{ \log \frac{P_1^{\otimes t}(x)}{P_0^{\otimes t}(x)} \right\} \\
\mathbbm{1} \left\{ \sum_{\tau=1}^t \sum_{j=1}^n \frac{\left[ W^\tau \right]_{ij}}{\pi_j} \log \frac{P_{j,1}(x_j^{(\tau)})}{P_{j,0}(x_j^{(\tau)})} \leq \gamma_i^{(t)} \right\}
\end{array}\right\} \nonumber \\
&= \sum_{x \in \mathcal{X}^t} P_0^{\otimes t}(x) \exp \left\{ h^{(t)}(x) \right\} \mathbbm{1} \left\{ \tilde{h}_i^{(t)}(x) \leq \gamma_i^{(t)} \right\} \nonumber \\
&= \sum_{x \in \mathcal{X}^t} P_0^{\otimes t}(x) \exp \left\{ \tilde{h}^{(t)}_i(x) + \varepsilon_i^{(t)}(x) \right\} \mathbbm{1} \left\{ \tilde{h}_i^{(t)}(x) \leq \gamma_i^{(t)} \right\}, \label{eq:thm1_beta_1}
\end{align}
where we let $\varepsilon_i^{(t)}(x) = h^{(t)}(x) - \tilde{h}^{(t)}_i(x)$. 

First, we deal with the term, $\varepsilon_i^{(t)}(x)$, with a convergence result on Markov chains.
\begin{lemma} \label{lem:markov_convergence}
Let $W$ be the transition matrix of some reversible, irreducible and aperiodic Markov chain, and let $1 = \lambda_1 > \lambda_2 \geq \dots \geq \lambda_n > -1$. Then, for all $i \in [n], t \in \mathbb{N}$ and $r_i \geq 0$,
\begin{align*}
    \left( \sum_{j=1}^n \left\vert [W^t]_{ij} - \pi_j \right\vert r_j \right)^2
    \leq \left( \frac{\pi_i}{1-\pi_i} \right) \left( \sum_{j=1}^n \pi_j r_j^2 \right) \rho^{2t}
\end{align*}
where $\rho = \max \left\{ \lambda_2 , \lvert \lambda_n \rvert \right\}$.
\end{lemma}
\begin{IEEEproof}[Proof of Lemma~\ref{lem:markov_convergence}]
The proof is similar to the one for Proposition~3 in~\cite{diaconis1991geometric}. We have
\begin{align*}
&\left( \sum_{j=1}^n \left\vert \left[ W^\tau \right]_{ij} - \pi_j \right\vert r_j \right)^2\\
&\leq \left( \sum_{j=1}^n \frac{1}{\pi_j} \left\vert \left[ W^\tau \right]_{ij} - \pi_j \right\vert^2 \right) \left( \sum_{j=1}^n \pi_j r_j^2 \right) \tag{Cauchy-Schwarz}
\end{align*}
and
\begin{align*}
&\sum_{j=1}^n \frac{1}{\pi_j} \left\vert \left[ W^\tau \right]_{ij} - \pi_j \right\vert^2\\
&= \sum_{j=1}^n \left( \frac{1}{\pi_j} \left( \left[ W^\tau \right]_{ij} \right)^2 - 2 \left[ W^\tau \right]_{ij} + \pi_j \right) \\
&= \left( \sum_{j=1}^n \frac{1}{\pi_i} \left[ W^\tau \right]_{ji} \left[ W^\tau \right]_{ij} \right) - 1 \tag{reversibility}
= \frac{1}{\pi_i} \left[ W^{2\tau} \right]_{ii} - 1.
\end{align*}
Following the remaining part in the proof for Proposition~3 in~\cite{diaconis1991geometric}, Lemma~\ref{lem:markov_convergence} is proved.
\end{IEEEproof}

By Lemma~\ref{lem:markov_convergence}, we can bound $\varepsilon_i^{(t)}(x)$ as
\begin{align}
    \varepsilon_i^{(t)}(x)
    =&\; \sum_{\tau=1}^t \sum_{j=1}^n \left( 1 - \frac{\left[ W^\tau \right]_{ij}}{\pi_j} \right) \log \frac{ P_{j,1}(x_j^{(\tau)}) }{ P_{j,0}(x_j^{(\tau)}) } \nonumber \\
    \leq&\; \sum_{\tau=1}^t \sum_{j=1}^n \left\vert \left[ W^\tau \right]_{ij} - \pi_j \right\vert \left\vert \frac{1}{\pi_j} \log \frac{ P_{j,1}(x_j^{(\tau)}) }{ P_{j,0}(x_j^{(\tau)}) } \right\vert \nonumber \\
    \leq&\; \sum_{\tau=1}^t \rho^\tau \sqrt{ \left( \frac{\pi_i}{1-\pi_i} \right) \sum_{j=1}^n \frac{1}{\pi_j} L_j^2 } \nonumber \\
    \leq&\; \frac{\rho}{1-\rho} \sqrt{ \left( \frac{\pi_i}{1-\pi_i} \right) \sum_{j=1}^n \frac{1}{\pi_j} L_j^2 } \label{eq:thm1_varepsilon}
\end{align}
for all $x \in \mathcal{X}^t$. 
Let
\begin{align*}
\tilde{S}_n^{(t)} &= \left( \frac{1}{t} \sum_{\tau=1}^t \sum_{j=1}^n \mathsf{Var}_{P_0} \left[ \frac{\left[ W^\tau \right]_{ij}}{\pi_j} \log \frac{ P_{j,1}(x_j) }{ P_{j,0}(x_j) } \right] \right)^{\frac{1}{2}},\\
y^{(t)} &= \frac{\tilde{h}_i^{(t)}(x) - \tilde{H}_i^{(t)} t}{ \tilde{S}_n^{(t)} \sqrt{t} }, \ 
\lambda^{(t)} = \frac{\gamma_i^{(t)} - \tilde{H}_i^{(t)} t}{ \tilde{S}_n^{(t)} \sqrt{t} }, 
\end{align*}
and plug~\eqref{eq:thm1_varepsilon} into~\eqref{eq:thm1_beta_1}, then we have
\begin{align}
&\beta^{(t)}_i(r^*; \eta^{(t)}_i, \gamma^{(t)}_i) \nonumber \\
&\leq \sum_{x \in \mathcal{X}^t} P_0^{\otimes t}(x) \exp \left\{ \tilde{h}^{(t)}_i(x) + \varepsilon_i^{(t)}(x) \right\} \mathbbm{1} \left\{ y^{(t)} \leq \lambda^{(t)} \right\} \nonumber \\
&\leq \exp \left\{ \frac{\rho}{1-\rho} \sqrt{ \left( \frac{\pi_i}{1-\pi_i} \right) \sum_{j=1}^n \frac{1}{\pi_j} L_j^2 } \right\} \notag \\
&\quad \cdot\sum_{y^{(t)}} P_{Y^{(t)}}(y^{(t)}) e^{y^{(t)} \tilde{S}_n^{(t)} \sqrt{t} + \tilde{H}_i^{(t)} t} \mathbbm{1} \left\{ y^{(t)} \leq \lambda^{(t)} \right\} \nonumber \\
&= e^{\tilde{H}_i^{(t)} t + \lambda^{(t)} \tilde{S}_n^{(t)} \sqrt{t} + \frac{\rho}{1-\rho} \sqrt{ \left( \frac{\pi_i}{1-\pi_i} \right) \sum_{j=1}^n \frac{1}{\pi_j} L_j^2 }} \notag\\ 
&\quad \cdot \sum_{y^{(t)} \leq \lambda^{(t)}} P_{Y^{(t)}}(y^{(t)}) \exp \Big\{ \underbrace{ \left( y^{(t)} - \lambda^{(t)} \right) \tilde{S}_n^{(t)} \sqrt{t} }_{z} \Big\} \nonumber \\
&= e^{\tilde{H}_i^{(t)} t + \lambda^{(t)} \tilde{S}_n^{(t)} \sqrt{t} + \frac{\rho}{1-\rho} \sqrt{ \left( \frac{\pi_i}{1-\pi_i} \right) \sum_{j=1}^n \frac{1}{\pi_j} L_j^2 }} \notag\\ 
&\quad \cdot \sum_{z \leq 0} P_{Y^{(t)}} \left( \frac{z}{\tilde{S}_n^{(t)} \sqrt{t}} + \lambda^{(t)} \right) e^z \nonumber \\
&=e^{\tilde{H}_i^{(t)} t + \lambda^{(t)} \tilde{S}_n^{(t)} \sqrt{t} + \frac{\rho}{1-\rho} \sqrt{ \left( \frac{\pi_i}{1-\pi_i} \right) \sum_{j=1}^n \frac{1}{\pi_j} L_j^2 }} \notag\\
&\quad\cdot \int_{z \leq 0} e^z d F_{Y^{(t)}} \left( \frac{z}{\tilde{S}_n^{(t)} \sqrt{t}} + \lambda^{(t)} \right). \label{eq:thm1_beta_2}
\end{align}

We further bound $\tilde{H}_i^{(t)}, \lambda^{(t)}, \tilde{S}_n^{(t)}$ and the integral term in~\eqref{eq:thm1_beta_2} individually. First, using Lemma~\ref{lem:markov_convergence},
\begin{align}
\tilde{H}_i^{(t)} 
&= \frac{1}{t} \mathsf{E}_{P_0} \left[ \tilde{h}^{(t)}(X) \right] \nonumber \\
&= \frac{1}{t} \mathsf{E}_{P_0} \left[ h^{(t)}_i(X) - \varepsilon^{(t)}_i(X) \right] \nonumber \\
&= H_n - \frac{1}{t} \sum_{\tau=1}^t \sum_{j=1}^n \left( 1 - \frac{\left[ W^\tau \right]_{ij}}{\pi_j} \right) \mathrm{D} \left( P_{j,0} \Vert P_{j,1} \right) \nonumber \\
&\leq H_n + \frac{1}{t} \frac{\rho}{1-\rho} \sqrt{ \left( \frac{\pi_i}{1-\pi_i} \right) \sum_{j=1}^n \frac{1}{\pi_j} \left( \mathrm{D} \left( P_{j,0} \Vert P_{j,1} \right) \right)^2 }. \label{eq:thm1_H}
\end{align}
For $\lambda^{(t)}$, we introduce a lemma from~\cite{esseen1945fourier}.
\begin{lemma}
\label{lem:esseen}
For i.i.d. random variables $X_1, X_2, \dots, X_n$ with non-lattice distributions, let $\sigma^2, \alpha_3$ denote the second and third central moment of each $X_i$. Let $F_n(\cdot)$ denote the CDF of $S_n \triangleq \frac{1}{\sigma} \sum_{i=1}^n X_i$, then
\begin{align*}
F_n(x) = \Phi(x) + \frac{\alpha_3}{6 \sqrt{2 \pi n} \sigma^3} e^{-\frac{x^2}{2}} (1-x^2) + o \left( \frac{1}{\sqrt{n}} \right),
\end{align*}
where $\Phi(\cdot)$ is the CDF of the standard normal distribution.
\end{lemma}
Following the proofs in~\cite{esseen1945fourier} and~\cite{cramer2004random}, the above Lemma~\ref{lem:esseen} could be extended to our case since each $\frac{[W^\tau]_{ij}}{\pi_j} \log \frac{P_{j,1}(X^{(\tau)}_j)}{P_{j,1}(X^{(\tau)}_j)}$ has its second and third moments in approximately the same scale with respect to $t$. In our case, the term $\sigma$ becomes the square of the mean of the $tn$ variance terms and $\alpha_3$ is the mean of the $tn$ third moment terms.

Now, let
\begin{align*}
\Delta \lambda^{(t)} &= \lambda^{(t)} - \lambda, \\
\Delta \Phi^{(t)} &= \Phi(\lambda^{(t)}) - \Phi(\lambda) = \Phi(\lambda^{(t)}) - (1-\epsilon).
\end{align*}
Since $F_{Y^{(t)}}((\lambda^{(t)})^-) < \Phi(\lambda) \leq F_{Y^{(t)}}(\lambda^{(t)})$, by Lemma~\ref{lem:esseen},
\begin{align*}
\Delta \Phi^{(t)}
&= \Phi(\lambda^{(t)}) - \Phi(\lambda^{(t)})\\
&\quad - \frac{\tilde{\alpha}_n}{6 \sqrt{2 \pi t} (\tilde{S}_n^{(t)})^3} \left( 1-(\lambda^{(t)})^2 \right) e^{-\frac{1}{2} (\lambda^{(t)})^2}\\
&\quad + o \left( 1/\sqrt{n} \right),
\end{align*}
where $(\tilde{S}_n^{(t)})^2 t$ and $\tilde{\alpha}_n t$ are the second and third moments of
\[
\sum_{\tau=1}^t \sum_{j=1}^n \frac{\left[ W^\tau \right]_{ij}}{\pi_j} \log \frac{ P_{j,1}(X_j) }{ P_{j,0}(X_j) }
\]
respectively. 
If we let $\varphi_n(x) = \frac{\tilde{\alpha}_n}{6 \tilde{S}_n^2} (1-x)^2$, we have
\begin{align}
\Delta \Phi^{(t)}
=&\; - \frac{1}{\tilde{S}_n^{(t)} \sqrt{t}} \Phi'(\lambda^{(t)}) \varphi_n(\lambda^{(t)}) + o \left( \frac{1}{\sqrt{t}} \right) \nonumber \\
=&\; - \frac{1}{\tilde{S}_n^{(t)} \sqrt{t}} \Phi'(\lambda) \varphi_n(\lambda) + o \left( \frac{1}{\sqrt{t}} \right) \label{eq:thm1_lambda1}
\end{align}
due to the fact that $\lim_{t \rightarrow \infty} \lambda^{(t)} = \lambda$ and the continuity of the functions. 

On the other hand,
\begin{align}
\Delta \Phi^{(t)}
&= \Phi'(\lambda) \Delta \lambda^{(t)} + o(\Delta \lambda^{(t)})
= \Phi'(\lambda) \Delta \lambda^{(t)} + o(\Delta \Phi^{(t)}) \notag \\
&= \Phi'(\lambda) \Delta \lambda^{(t)} + O \left( \frac{1}{\sqrt{t}} \right), \label{eq:thm1_lambda2}
\end{align}
and from~\eqref{eq:thm1_lambda1}, 
\begin{align*}
\Delta \Phi^{(t)} = O \left( \frac{1}{\sqrt{t}} \right).
\end{align*}
From~\eqref{eq:thm1_lambda1} and~\eqref{eq:thm1_lambda2} we have
\begin{align}
\lambda^{(t)}
= \lambda + \Delta \lambda^{(t)}
= \lambda - \frac{1}{\tilde{S}_n^{(t)} \sqrt{t}} \varphi_n(\lambda) + o \left( \frac{1}{\sqrt{t}} \right). \label{eq:thm1_lambda}
\end{align}
For $\tilde{S}_n^{(t)}$, by Lemma~\ref{lem:markov_convergence},
\begin{align*}
&(\tilde{S}_n^{(t)})^2 - S_n^2\\
&= \frac{1}{t} \sum_{\tau=1}^t \sum_{j=1}^n \left( \left( \frac{\left[ W^\tau \right]_{ij}}{\pi_j} \right)^2 - 1 \right) \mathrm{Var}_{P_0} \left[ \log \frac{ P_{j,1}(X_j^{(\tau)}) }{ P_{j,0}(X_j^{(\tau)}) } \right] \\
&= \frac{1}{t} \sum_{\tau=1}^t \sum_{j=1}^n \left( \frac{\left[ W^\tau \right]_{ij}}{\pi_j} - 1 \right) \left( \frac{\left[ W^\tau \right]_{ij}}{\pi_j} + 1 \right) \sigma^2_{j} \\
&\leq \frac{1}{t} \left( \frac{1+\pi_{\min}}{\pi_{\min}} \right) \sum_{\tau=1}^t \sum_{j=1}^n \left\vert \frac{\left[ W^\tau \right]_{ij}}{\pi_j} - 1 \right\vert \sigma^2_{j} \\
&\leq \frac{1}{t} \left( \frac{1+\pi_{\min}}{\pi_{\min}} \right) \left( \frac{\rho}{1-\rho} \right) \sqrt{ \left( \frac{\pi_i}{1-\pi_i} \right) \sum_{j=1}^n \frac{1}{\pi_j} \sigma^4_j }
\end{align*}
where $\pi_{\min} = \min_{i \in [n]} \pi_i$ and $\sigma^2_j = \mathrm{Var}_{P_0} \left[ \log \frac{ P_{j,1}(X_j^{(\tau)}) }{ P_{j,0}(X_j^{(\tau)}) } \right]$. Thus, we have
\begin{align*}
&\tilde{S}_n^{(t)} - S_n\\
&\leq \frac{\frac{1}{t} \left( \frac{1+\pi_{\min}}{\pi_{\min}} \right) \left( \frac{\rho}{1-\rho} \right) }{S_n + \tilde{S}_n^{(t)}}  \sqrt{ \left( \frac{\pi_i}{1-\pi_i} \right) \sum_{j=1}^n \frac{1}{\pi_j} \sigma^4_j } \\
&= \left( \frac{1}{2 S_n} + o(1) \right) \frac{\left( \frac{1+\pi_{\min}}{\pi_{\min}} \right) \left( \frac{\rho}{1-\rho} \right)}{t}  \sqrt{ \left( \frac{\pi_i}{1-\pi_i} \right) \sum_{j=1}^n \frac{1}{\pi_j} \sigma^4_j }
\end{align*}
and
\begin{align}
\tilde{S}_n^{(t)} = S_n + O \left( \frac{1}{t} \right). \label{eq:thm1_S}
\end{align}
Plug~\eqref{eq:thm1_H},~\eqref{eq:thm1_lambda} and~\eqref{eq:thm1_S} into~\eqref{eq:thm1_beta_2}, then we have
\begin{align}
&\beta_i^{(t)}(r^*; \eta^{(t)}_i, \gamma^{(t)}_i) \notag\\
&\leq \exp \left\{ H_n t + \lambda S_n \sqrt{t} - \varphi_n(\lambda) \right\} \notag\\
&\quad\cdot\exp \left\{ \frac{\rho}{1-\rho} \sqrt{\frac{\pi_i}{1-\pi_i}} \sqrt{ \sum_{j=1}^n \frac{1}{\pi_j} \left( \mathrm{D} \left( P_{j,0} \Vert P_{j,1} \right) \right)^2 } \right\} \nonumber \\
&\quad\cdot \exp \left\{ \frac{\rho}{1-\rho} \sqrt{\frac{\pi_i}{1-\pi_i}} \sqrt{ \sum_{j=1}^n \frac{1}{\pi_j} L_j^2 } \right\} \cdot A \label{eq:thm1_beta_3}
\end{align}
where
\begin{align*}
A = \int_{z \leq 0} e^z d F_{Y^{(t)}} \left( \frac{z}{\tilde{S}_n^{(t)} \sqrt{t}} + \lambda^{(t)} \right).
\end{align*}

We then deal with the remaining integral term. 
Let
\begin{align*}
B(x) = \frac{\tilde{\alpha}_n}{b \sqrt{2 \pi t} (\tilde{S}_n^{(t)})^3} \left( 1 - x^2 \right) e^{-\frac{x^2}{2}}
\end{align*}
and again, by Lemma~\ref{lem:esseen}
\begin{align}
&\int_{z \leq 0} e^z d F_{Y^{(t)}} \left( \frac{z}{\tilde{S}_n^{(t)}} + \lambda^{(t)} \right) \nonumber \\
&=\int_{z \leq 0} e^z d \Phi \left( \frac{z}{\tilde{S}_n^{(t)} \sqrt{t}} + \lambda^{(t)} \right) \notag\\
&\quad + \frac{1}{\sqrt{t}} \int_{z \leq 0} e^z d B \left( \frac{z}{\tilde{S}_n^{(t)} \sqrt{t}} + \lambda^{(t)} \right) + o \left( \frac{1}{\sqrt{t}} \right) \nonumber \\
&= \frac{1}{\tilde{S}_n^{(t)} \sqrt{t}} \int_{z \leq 0} \frac{1}{\sqrt{2 \pi}} e^{z - \frac{1}{2} \left( \frac{z}{\tilde{S}_n^{(t)} \sqrt{t}} + \lambda^{(t)} \right)^2} dz + \frac{1}{\sqrt{t}} B(\lambda^{(t)}) \notag\\
&\quad - \frac{1}{\sqrt{t}} \int_{z \leq 0} e^z B \left( \frac{z}{\tilde{S}_n^{(t)} \sqrt{t}} + \lambda^{(t)} \right) dz + o \left( \frac{1}{\sqrt{t}} \right) \nonumber \\
&= \left( \frac{1}{S_n} + o(1) \right) \frac{1}{\sqrt{t}} \int_{z \leq 0} e^{z - \frac{\lambda}{2} + o(1)} + \frac{1}{\sqrt{t}} B(\lambda) \notag\\
&\quad - \frac{1}{\sqrt{t}} \int_{z \leq 0} e^z \left( B(\lambda) + o(1) \right) dz + o \left( \frac{1}{\sqrt{t}} \right) \nonumber \\
&= \frac{1}{S_n \sqrt{2 \pi t}} e^{-\frac{\lambda^2}{2}} + o \left( \frac{1}{\sqrt{t}} \right) \nonumber \\
&= \frac{1}{S_n \sqrt{2 \pi t}} e^{-\frac{\lambda^2}{2} + o(1)} \label{eq:thm1_int}.
\end{align}
Plug~\eqref{eq:thm1_int} into~\eqref{eq:thm1_beta_3}, we have
\begin{align*}
    \beta_i^{*(t)}(r^*;\epsilon)
    \leq \beta_i^{*(t)}(r^*; \eta^{(t)}_i, \gamma^{(t)}_i)
    \leq C^{(\mathrm{NP})}_i \beta^{(t)*}_{\mathrm{cen}}(\epsilon),
\end{align*}
where the optimal type-II error probability could be found in~\cite{strassen1962asymptotische} as
\begin{align*}
    \beta^{(t)*}_{\mathrm{cen}}(\epsilon)
    = e^{H_n t + \lambda S_n \sqrt{t} - \frac{1}{2} \log t - \frac{1}{2} \log \left( 2\pi  \right) - \log S_n - \varphi_n(\lambda) -\frac{\lambda^2}{2}}.
\end{align*}

\subsection{Proof of Theorem~\ref{thm:bayes_higher_order}}
\label{pf:bayes_higher_order}
We know that for the centralized Bayes setting with prior $\xi=(\xi_0,\xi_1)$, a log-likelihood ratio test with threshold $\eta \coloneqq \log \frac{\xi_0}{\xi_1}$ minimizes the Bayes risk.
For the decentralized case, though the optimal threshold may not be $\eta$, we use the threshold for testing and view the induced Bayes risk as an upper bound on the optimal Bayes risk. \\
First, let us consider the exponentially tilted distribution for each node $i$,
\begin{align*}
P_{i,\theta}(x) \propto \left( P_{i,0}(x) \right)^{1-\theta} \left( P_{i,1}(x) \right)^{\theta} \quad \forall x \in \mathcal{X}_i.
\end{align*}
For the product distribution of the nodes, first let $x$ denote $(x^{(1)}_1, x{(t)}_2, \dots, x^{(t)}_n) \in \mathcal{X}^t$, and we have
\begin{align*}
P_\theta(x) \propto \left( P_0(x) \right)^{1-\theta} \left( P_1(x) \right)^{\theta} \quad \forall x \in \mathcal{X}.
\end{align*}
Furthermore, let $\theta^* \in [0,1]$ be
\begin{align*}
\theta^* = \underset{\theta \in [0,1]}{\arg \max} - \log \mathsf{E}_{X \sim P_0} \left[ \left( \frac{P_1(X)}{P_0(X)} \right)^{\theta} \right],
\end{align*}
which means that we have
\begin{align*}
D_{\mathrm{KL}} \left( P_{\theta^*} \Vert P_0 \right) = D_{\mathrm{KL}} \left( P_{\theta^*} \Vert P_1 \right) = \mathrm{CI} \left( P_0, P_1 \right).
\end{align*}
For the type-II error, we have
\begin{align}
&\beta^{(t)}_i(\xi;\eta) \notag\\
&= P_1 \left\{ \sum_{\tau=1}^t \sum_{j=1}^n \frac{[W^\tau]_{ij}}{\pi_j} \log \frac{P_{j,1}(X^{(t)}_j)}{P_{j,0}(X^{(t)}_j)} \leq \eta \right\} \nonumber \\
&= \sum_{x \in \mathcal{X}^t} P^{\otimes t}_1(x) \mathbbm{1} \left\{ \sum_{\tau=1}^t \sum_{j=1}^n \frac{[W^\tau]_{ij}}{\pi_j} \log \frac{P_{j,1}(x^{(t)}_j)}{P_{j,0}(x^{(t)}_j)} \leq \eta \right\}. \label{eq:bayes_higher_0}
\end{align}
Applying the change of measure from $P_1$ to $P_{\theta^*}$, we have
\begin{align}
\eqref{eq:bayes_higher_0}
&= \sum_{x \in \mathcal{X}^t} P^{\otimes t}_{\theta^*}(x) \exp \left\{ \sum_{\tau=1}^t \sum_{j=1}^n \log \frac{P_{j,1}(x^{(t)}_j)}{P_{j,\theta}(x^{(t)}_j)} \right\} \notag\\
&\quad \cdot\mathbbm{1} \left\{ \sum_{\tau=1}^t \sum_{j=1}^n \frac{[W^\tau]_{ij}}{\pi_j} \log \frac{P_{j,1}(x^{(t)}_j)}{P_{j,0}(x^{(t)}_j)} \leq \eta \right\}. \label{eq:bayes_higher_change_of_measure}
\end{align}
By the definition of the tilted distribution $P_{\theta^*}$, we have the following equality:
\begin{align*}
&\sum_{j=1}^n \log \frac{P_{j,1}(x^{(t)}_j)}{P_{j,\theta}(x^{(t)}_j)}\\
&= (1-\theta^*) \left( \sum_{j=1}^n \log \frac{P_{j,1}(x^{(t)}_j)}{P_{j,0}(x^{(t)}_j)} \right) + \mathrm{CI}(P_0, P_1).
\end{align*}
Let $\mathrm{CI} = \mathrm{CI}(P_0, P_1)$ and we have
\begin{align}
&\eqref{eq:bayes_higher_change_of_measure} \notag\\
&= \sum_{x \in \mathcal{X}^t} 
\left\{
\begin{array}{l} P^{\otimes t}_{\theta^*}(x) 
e^{(1-\theta^*) \left( \sum_{\tau=1}^t \sum_{j=1}^n \log \frac{P_{j,0}(x^{(t)}_j)}{P_{j,\theta}(x^{(t)}_j)} \right) - \mathrm{CI} t} \\
\cdot\mathbbm{1} \left\{ \sum_{\tau=1}^t \sum_{j=1}^n \frac{[W^\tau]_{ij}}{\pi_j} \log \frac{P_{j,1}(x^{(t)}_j)}{P_{j,0}(x^{(t)}_j)} \leq \eta \right\}
\end{array}\right\}. \label{eq:bayes_higher_CI}
\end{align}
Similar to the trick in Appendix~\ref{pf:neyman_pearson_higher_order}, we have
\begin{align*}
&\sum_{\tau=1}^t \sum_{j=1}^n (1-\theta^*) \log \frac{P_{j,1}(x^{(t)}_j)}{P_{j,0}(x^{(t)}_j)} \\
&= \sum_{\tau=1}^t \sum_{j=1}^n (1-\theta^*) \frac{[W^\tau]_{ij}}{\pi_j} \log \frac{P_{j,1}(x^{(t)}_j)}{P_{j,0}(x^{(t)}_j)}\\
&\quad + \sum_{\tau=1}^t \sum_{j=1}^n \left( 1 - (1-\theta^*) \frac{[W^\tau]_{ij}}{\pi_j} \right) \log \frac{P_{j,1}(x^{(t)}_j)}{P_{j,0}(x^{(t)}_j)} \\
&\leq \sum_{\tau=1}^t \sum_{j=1}^n (1-\theta^*) \frac{[W^\tau]_{ij}}{\pi_j} \log \frac{P_{j,1}(x^{(t)}_j)}{P_{j,0}(x^{(t)}_j)}\\
&\quad + \underbrace{(1-\theta^*) \frac{\rho}{1-\rho} \sqrt{ \frac{1-\pi_i}{\pi_i} \left( \sum_{j=1}^n \frac{1}{\pi_j} L^2_j \right) }}_{C_i}.
\end{align*}
Let $w(x)$ denote the weighted sum of the log-likelihood ratios, then we have
\begin{align}
&\eqref{eq:bayes_higher_CI} \notag\\
&= e^{ -\mathrm{CI} t + (1-\theta^*) C_i }\! \sum_{x \in \mathcal{X}^t} P^{\otimes t}_{\theta^*}(x) 
e^{ (1-\theta^*) w(x)} \mathbbm{1} \left\{ w(x) \leq \eta\right\}. \label{eq:bayes_higher_w(x)}
\end{align}
To invoke Esseen's theorem~\cite{esseen1945fourier}, let
\begin{align*}
\sigma_j^2 &= \mathsf{Var}_{P_{j, \theta^*}} \left[ \log \frac{P_{j,1}(X_j)}{P_{j,0}(X_j)} \right], \\
S^{(t)}_n &= \left[ \frac{1}{t} \sum_{\tau=1}^t \sum_{j=1}^n \left( \frac{[W^\tau]_{ij}}{\pi_j} \right)^2 \sigma^2_j \right]^{\frac{1}{2}},
\end{align*}
and let $Y^{(t)}_i$ denote the normalized tilted sum of the log-likelihood ratios
\begin{align*}
Y^{(t)}_i = \frac{1}{S^{(t)}_n \sqrt{t}} \left( \sum_{\tau=1}^t \sum_{j=1}^n (1-\theta^*) \frac{[W^\tau]_{ij}}{\pi_j} \log \frac{P_{j,1}(X^{(t)}_j)}{P_{j,0}(X^{(t)}_j)} \right).
\end{align*}
By Lemma~\ref{lem:esseen}, we know that $Y^{(t)}_i$'s CDF converges to the one of the standard normal distribution with some remaining terms, and so far we have equation~\eqref{eq:bayes_higher_w(x)} become
\begin{align*}
&e^{- \mathrm{CI} t + (1-\theta^*) C_i} \\
&\quad\cdot \sum_{y \in \mathcal{Y}} 
\Bigg(\begin{array}{l}
P_{Y^{(t)}_i}(y)e^{ (1-\theta^*) S^{(t)}_n \sqrt{t} y}
\\\mathbbm{1} \left\{ (1-\theta^*) S^{(t)}_n \sqrt{t} yx \leq (1-\theta^*) \eta \right\}
\end{array}\Bigg)
\\
&= e^{- \mathrm{CI} t + (1-\theta^*) \eta +  (1-\theta^*) C_i}  \sum_{z \leq 0} P_{Y^{(t)}_i} \left( \frac{z+(1-\theta^*)\eta}{(1-\theta^*) S^{(t)}_n \sqrt{t}} \right) e^z.
\end{align*}
The summation term could be
\begin{align*}
\sum_{z \leq 0} P_{Y^{(t)}_i} \left( \frac{z+(1-\theta^*)\eta}{(1-\theta^*) S^{(t)}_n \sqrt{t}} \right) e^z
= \frac{1}{(1-\theta^*) \sigma \sqrt{2 \pi t}} e^{o(1)}
\end{align*}
using the similar method in Appendix~\ref{pf:neyman_pearson_higher_order}.
Thus, the type-II error probability is
\begin{align}
&\beta^{(t)}_i(\xi;\eta) \notag\\
&=\exp\left\{
\begin{array}{l}
- \mathrm{CI} t - \frac{1}{2} \log t + (1-\theta^*) \eta + (1-\theta^*) C_i\\
 - \log \left( (1-\theta^*) \sigma \right) - \frac{1}{2} \log (2\pi) + o(1)
\end{array}
\right\}. \label{eq:bayes_higher_beta}
\end{align}
Similarly, the type-I error probability is
\begin{align}
&\alpha^{(t)}_i(\xi;\eta) \notag\\ 
&= \exp \left\{ 
\begin{array}{l}
- \mathrm{CI} t - \frac{1}{2} \log t - \theta^* \eta + \theta^* C_i\\
 - \log \left( \theta^* \sigma \right) - \frac{1}{2} \log (2\pi) + o(1) 
\end{array}\right\}. \label{eq:bayes_higher_alpha}
\end{align}
Let $\bar{\theta} = \max \left\{ \theta^*, 1-\theta^* \right\} = \frac{1}{2} + \lvert \theta^* - \frac{1}{2} \rvert$ and
\begin{align*}
\bar{\alpha}^{(t)}_i(\xi;\eta)
&= \exp\left\{
\begin{array}{l}
- \mathrm{CI} t - \frac{1}{2} \log t - \theta^* \eta + \bar{\theta} C_i\\
 - \log \left( \theta^* \sigma \right) - \frac{1}{2} \log (2\pi) + o(1)
 \end{array}\right\}, \\
\bar{\beta}^{(t)}_i(\xi;\eta)
&= \exp\left\{
\begin{array}{l}
- \mathrm{CI} t - \frac{1}{2} \log t + (1-\theta^*) \eta + \bar{\theta} C_i\\
 - \log \left( (1-\theta^*) \sigma \right) - \frac{1}{2} \log (2\pi) + o(1)
 \end{array}\right\}.
\end{align*}
Since $\eta = \log \frac{\xi_0}{\xi_1}$, from~\eqref{eq:bayes_higher_beta} and~\eqref{eq:bayes_higher_alpha}, we have
\begin{align*}
&\mathsf{P}_{\mathsf{e},i}^{(t)*}(\xi)\\
&\leq \xi_0 \alpha^{(t)}_i(\xi;\eta) + \xi_1 \beta^{(t)}_i(\xi;\eta)\\
&\leq  \xi_0 \bar{\alpha}^{(t)}_i(\xi;\eta) + \xi_1 \bar{\beta}^{(t)}_i(\xi;\eta) \\
&= \bar{\alpha}^{(t)}_i(\xi;\eta) \left( \xi_0 + \xi_1 \left( \frac{\bar{\beta}^{(t)}_i(\xi;\eta)}{\bar{\alpha}^{(t)}_i(\xi;\eta)} \right) \right) \\
&= \bar{\alpha}^{(t)}_i(\xi;\eta) \left( \xi_0 + \xi_1 \left( \frac{\xi_0}{\xi_1} \right) \left( \frac{\theta^*}{1-\theta^*} \right) \right) \\
&= \xi_0^{1-\theta^*} \xi_1^{\theta^*} \exp \left\{ 
\begin{array}{l}
- \mathrm{CI} t - \frac{1}{2} \log t + \bar{\theta} C_i\\
 - \log \left( \theta^* (1-\theta^*) \sigma \right) - \frac{1}{2} \log (2\pi)\\ 
 + o(1) 
 \end{array}\right\} \\
&= e^{\bar{\theta} C_i}  \mathsf{P}_{\mathsf{e},\mathrm{cen}}^{(t)*}(\xi) \\
&= C^{(\mathrm{B})}_i \mathsf{P}_{\mathsf{e},\mathrm{cen}}^{(t)*}(\xi),
\end{align*}
where $\mathsf{P}_{\mathsf{e},\mathrm{cen}}^{(t)*}(\xi)$ is the optimal Bayes risk in the centralized case which could be found in~\cite{efron1968large}.

\subsection{Gaussian Case}
\label{pf:NP_gaussian}
For a network with $n$ nodes, let the observations follow the Gaussian distribution such that
\begin{align*}
    \mathcal{H}_0:&\; \;X^{(t)}_i \overset{\text{i.i.d.}}{\sim} \mathrm{Normal}(-\mu, \sigma^2), \\
    \mathcal{H}_1:&\; \;X^{(t)}_i \overset{\text{i.i.d.}}{\sim} \mathrm{Normal}(\mu, \sigma^2),
\end{align*}
for all $i \in [n], t \in \mathbb{N}$. Then, the log-likelihood ratio is
\begin{align*}
    \log \frac{P_{i,1}(X_i)}{P_{i,0}(X_i)}
    =&\; \log \frac{ \frac{1}{\sqrt{2\pi \sigma^2}} e^{\frac{-(X_i-\mu)^2}{2\sigma^2}} }{ \frac{1}{\sqrt{2\pi \sigma^2}} e^{\frac{-(X_i+\mu)^2}{2\sigma^2}} }
    = \frac{2\mu}{\sigma} X_i
\end{align*}
for all $i \in [n]$. 

At each time $t$, node $i$ has its log-belief ratio as
\begin{align*}
    \sum_{\tau=1}^t \sum_{j=1}^n \frac{[W^\tau]_{ij}}{\pi_j} \log \frac{P_{j,1}(X^{(t-\tau+1)}_j)}{P_{j,1}(X^{(t-\tau+1)}_j)}
\end{align*}
which follows the distribution
\begin{align*}
    \mathcal{H}_0:&\; \;\mathrm{Normal}(-\tilde{\mu}, \tilde{\sigma}^2), \\
    \mathcal{H}_1:&\; \;\mathrm{Normal}(\tilde{\mu}, \tilde{\sigma}^2)
\end{align*}
with
\begin{align*}
    \tilde{\mu} &= \left( \sum_{\tau,j=1}^{t,n} \frac{[W^\tau]_{ij}}{\pi_j} \right) \frac{2 \mu^2}{\sigma^2}, \\
    \tilde{\sigma}^2 &= \left( \sum_{\tau,j=1}^{t,n} \left( \frac{[W^\tau]_{ij}}{\pi_j} \right)^2 \right) \frac{4 \mu^2}{\sigma^2}.
\end{align*}
For the Neyman-Pearson problem, the optimal threshold, $\gamma^*$ is set such that the type-I error is $\epsilon$, that is
\begin{align*}
    \mathcal{Q} \left( \frac{\gamma^* + \tilde{\mu}}{\tilde{\sigma}} \right) = \epsilon
\end{align*}
where $\mathcal{Q}(\cdot)$ is the Q-function for the standard normal distribution, and thus
\begin{align*}
    \gamma^* = - \tilde{\mu} + \tilde{\sigma} \mathcal{Q}^{-1}(\epsilon).
\end{align*}
Let $\lambda = \mathcal{Q}^{-1}(\epsilon)$. Now the type-II error is
\begin{align}
    \Phi \left( \frac{\gamma^* - \tilde{\mu}}{\tilde{\sigma}} \right)
    =&\; \mathcal{Q} \left( \frac{- \gamma^* + \tilde{\mu}}{\tilde{\sigma}} \right)
    = \mathcal{Q} \left( \frac{2 \tilde{\mu}}{\tilde{\sigma}} - \lambda \right). \label{eq:gaussian_typeII}
\end{align}
Since we know that $\mathcal{Q}(x) < \frac{1}{x \sqrt{2\pi}} e^{\frac{-x^2}{2}}$, let $A=\frac{2 \tilde{\mu}}{\tilde{\sigma}} - \lambda$ we have
\begin{align}
    \eqref{eq:gaussian_typeII}
    &< \frac{1}{A\sqrt{2\pi}} \exp \left\{ \frac{-1}{2} \left( \frac{4\tilde{\mu}^2}{\tilde{\sigma}^2} - \frac{4\tilde{\mu}}{\tilde{\sigma}} \lambda + \lambda^2 \right) \right\} \nonumber \\
    &= \frac{1}{A\sqrt{2\pi}} \exp \left\{ \frac{-2\mu}{\sigma}B  + \frac{2\mu}{\sigma} C\lambda - \frac{\lambda^2}{2} \right\}, \label{eq:gaussian_typeII2}
\end{align}
where the shorthand notations 
\[
B = \frac{ \left( \sum_{\tau,j=1}^{t,n} \frac{[W^\tau]_{ij}}{\pi_j} \right)^2 }{ \sum_{\tau,j=1}^{t,n} \left( \frac{[W^\tau]_{ij}}{\pi_j} \right)^2 },\ 
C = \frac{ \left( \sum_{\tau,j=1}^{t,n} \frac{[W^\tau]_{ij}}{\pi_j} \right) }{ \sqrt{ \sum_{\tau,j=1}^{t,n} \left( \frac{[W^\tau]_{ij}}{\pi_j} \right)^2 } }.
\]
For the term $B$ in~\eqref{eq:gaussian_typeII2},
\begin{align*}
    B &= \frac{ \left[ tn + \sum_{\tau,j=1}^{t,n} \left( \frac{[W^\tau]_{ij}}{\pi_j} - 1 \right) \right]^2 }{ tn + \sum_{\tau,j=1}^{t,n} \left[ \left( \frac{[W^\tau]_{ij}}{\pi_j} \right)^2 - 1 \right] } \\
    &= \frac{ \left(
    \begin{array}{l} tn + 2 \sum_{\tau,j=1}^{t,n} \left( \frac{[W^\tau]_{ij}}{\pi_j} - 1 \right)\\
     + \frac{1}{tn} \left[ \sum_{\tau,j=1}^{t,n} \left( \frac{[W^\tau]_{ij}}{\pi_j} - 1 \right) \right]^2
     \end{array} \right)
     }{ 1 + \frac{1}{tn} \sum_{\tau,j=1}^{t,n} \left[ \left( \frac{[W^\tau]_{ij}}{\pi_j} \right)^2 - 1 \right] } \\
    &= \left[ tn + 2 \sum_{\tau,j=1}^{t,n} \left( \frac{[W^\tau]_{ij}}{\pi_j} - 1 \right) + o \left( \frac{1}{t} \right) \right] \\
    &\quad \cdot\left[ 1 - \frac{1}{tn} \sum_{\tau,j=1}^{t,n} \left[ \left( \frac{[W^\tau]_{ij}}{\pi_j} \right)^2 - 1 \right] + o \left( \frac{1}{t} \right) \right]\\
    &= \left[ tn + 2 \sum_{\tau,j=1}^{t,n} \left( \frac{[W^\tau]_{ij}}{\pi_j} - 1 \right) \right]\\
    &\quad\cdot \left[ \begin{array}{l}1 - \frac{2}{tn} \sum_{\tau,j=1}^{t,n} \left( \frac{[W^\tau]_{ij}}{\pi_j} - 1 \right)\\ - \frac{1}{tn} \sum_{\tau,j=1}^{t,n} \left( \frac{[W^\tau]_{ij}}{\pi_j} - 1 \right)^2 \end{array}\right] + o(1) \\
    &= tn - \sum_{\tau,j=1}^{t,n} \left( \frac{[W^\tau]_{ij}}{\pi_j} - 1 \right)^2 + o(1).
\end{align*}
By Lemma~\ref{lem:markov_convergence}, we can see that
\begin{align*}
    B &\geq tn - \sum_{\tau=1}^t \left( \frac{\pi_i}{1-\pi_i} \right) \left( \sum_{j=1}^n \frac{1}{\pi_j} \right) \rho^{2\tau} + o(1)\\
    &= tn - \left( \frac{\rho^2}{1-\rho^2} \right) \left( \frac{\pi_i}{1-\pi_i} \right) \left( \sum_{j=1}^n \frac{1}{\pi_j} \right) + o(1).
\end{align*}
For the term $C$ in~\eqref{eq:gaussian_typeII2}, we have
\begin{align*}
    C = B^{\frac{1}{2}} &= \sqrt{tn - \sum_{\tau,j=1}^{t,n} \left( \frac{[W^\tau]_{ij}}{\pi_j} - 1 \right)^2 + o(1)}\\
     &= \sqrt{tn} + o(1).
\end{align*}
Finally, for the term $A$ in~\eqref{eq:gaussian_typeII2}, from~\eqref{eq:gaussian_typeII2} we can see that
\begin{align*}
    A + \lambda = \frac{2\mu}{\sigma} C = \frac{2\mu\sqrt{n}}{\sigma} \sqrt{t} + o(1).
\end{align*}
Thus, we have
\begin{align*}
    A &= \left( \frac{2\mu\sqrt{n}}{\sigma} \sqrt{t} - \lambda + o(1) \right)\\
    &= \frac{1}{\frac{2\mu\sqrt{n}}{\sigma} \sqrt{t}}(1+o(1))
    =  \frac{1}{\frac{2\mu\sqrt{n}}{\sigma} \sqrt{t}} e^{o(1)}.
\end{align*}
Plug $A$, $B$, and $C$ back to~\eqref{eq:gaussian_typeII2}, we have
\begin{align*}
    &\beta^{(t)*}_i(r^*,\epsilon)\\
    &\leq \frac{1}{\frac{2\mu \sqrt{n}}{\sigma} \sqrt{2\pi t}} \exp \left\{\begin{array}{l}
     - \frac{2\mu n}{\sigma} t + \frac{2 \mu \lambda \sqrt{n}}{\sigma} \sqrt{t} - \frac{\lambda^2}{2}\\
      + \frac{2\mu}{\sigma} \big( \frac{\rho^2}{1-\rho^2} \big) \big( \frac{\pi_i}{1-\pi_i} \big) \!\sum\limits_{j=1}^n \frac{1}{\pi_j} \\ + o(1) 
      \end{array}\right\}.
\end{align*}
With Strassen's result in~\cite{strassen2009asymptotic}, we can show that the optimal type-II error in the centralized case is
\begin{align*}
    &\beta^{(t)*}_{\mathrm{cen}}(\epsilon)\\
    &= \frac{1}{\frac{2\mu \sqrt{n}}{\sigma} \sqrt{2\pi t}} \exp \left\{ - \frac{2\mu n}{\sigma} t + \frac{2 \mu \lambda \sqrt{n}}{\sigma} \sqrt{t} - \frac{\lambda^2}{2} + o(1) \right\}.
\end{align*}
Comparing the centralized and decentralized case, we can see that
\begin{align*}
    \beta^{(t)*}_i(r^*,\epsilon)
    \leq C^{(\mathrm{B,Gaussian})}_i \cdot \beta^{(t)*}_{\mathrm{cen}}(\epsilon)
\end{align*}
where
\begin{align*}
    C^{(\mathrm{B,Gaussian})}_i = \frac{2\mu}{\sigma} \left( \frac{\rho^2}{1-\rho^2} \right) \left( \frac{\pi_i}{1-\pi_i} \right) \left( \sum_{j=1}^n \frac{1}{\pi_j} \right).
\end{align*}

\bibliographystyle{IEEEtran}

\begin{IEEEbiographynophoto}{Bruce (Yu-Chieh) Huang}
(Student Member, IEEE) received the B.Sc. degree in Electrical Engineering from National Taiwan University, Taiwan, in 2019, and the M.S. degree in Communication Engineering from the same university in 2021. He is currently a pursuing a Ph.D. degree in Electrical and Computer Engineering at University of California, Los Angeles, USA. His research interests include information theory and statistical learning.
\end{IEEEbiographynophoto}

\begin{IEEEbiographynophoto}{I-Hsiang Wang}(Member, IEEE) received the B.Sc. degree in Electrical Engineering from National Taiwan University, Taiwan, in 2006. He received a Ph.D. degree in Electrical Engineering and Computer Sciences from the University of California at Berkeley, USA, in 2011. From 2011 to 2013, he was a postdoctoral researcher at \`{E}cole Polytechnique F\`{e}d\`{e}rale de Lausanne, Switzerland. Since 2013, he has been at the Department of Electrical Engineering in National Taiwan University, where he is now a professor. His research interests include network information theory, networked data analysis, and statistical learning. He was a finalist of the Best Student Paper Award of IEEE International Symposium on Information Theory, 2011. He received the 2017 IEEE Information Theory Society Taipei Chapter and IEEE Communications Society Taipei/Tainan Chapters Best Paper Award for Young Scholars.
\end{IEEEbiographynophoto}

\end{document}